\documentclass[twocolumn,epjc3]{svjour3}  
\smartqed  
\usepackage{fix-cm}
\usepackage{graphicx}
\usepackage{amsmath}
\usepackage{mathrsfs}
\usepackage{mathtools}
\usepackage{booktabs}
\usepackage{comment}
\usepackage[colorlinks,urlcolor=blue,linkcolor=blue,citecolor=blue]{hyperref}
\usepackage[ruled,linesnumbered]{algorithm2e}
\usepackage{setspace}
\usepackage{multirow}
\usepackage{multicol}
\usepackage{url}
\usepackage{tikz}
\usepackage{color,soul}
\usepackage{adjustbox}
\usepackage{graphicx}
\usepackage{makecell}
\usepackage{multirow}
\usepackage{enumitem}
\usepackage{microtype}
\usepackage{algorithmic}
\usepackage{textcomp}
\usepackage{array}
\usepackage{dblfloatfix}
\usepackage{subcaption}
\usepackage{pgfplots}
\usepackage{amssymb}
\usepackage{tabularray}
\usepgfplotslibrary{groupplots}
\usepackage{subcaption,siunitx,booktabs}

\DeclareMathOperator*{\argmin}{arg\,min}

\begin{document}

\title{Differentially Private GANs for Generating Synthetic Indoor Location Data}

\author{Vahideh Moghtadaiee\thanksref{*,addr1}
        \and
        Mina Alishahi\thanksref{addr2}
        \and
        Milad Rabiei\thanksref{addr3}
}

\thankstext{*}{e-mail: {v\_moghtadaiee@sbu.ac.ir}}


\institute{Cyberspace Research Institute, Shahid Beheshti University, Tehran, Iran \label{addr1}
           \and
           Department of Computer Science, Open Universiteit, The Netherlands \label{addr2}
           \and
           School of Electrical and Computer Engineering, Shahid Beheshti University, Tehran, Iran \label{addr3}
}


\maketitle

\begin{abstract}
The advent of location-based services has led to the widespread adoption of indoor localization systems, which enable location tracking of individuals within enclosed spaces such as buildings. While these systems provide numerous benefits such as improved security and personalized services, they also raise concerns regarding privacy violations. As such, there is a growing need for privacy-preserving solutions that can protect users' sensitive location information while still enabling the functionality of indoor localization systems.
In recent years, Differentially Private Generative Adversarial Networks (DPGANs) have emerged as a powerful methodology that aims to protect the privacy of individual data points while generating realistic synthetic data similar to original data.  DPGANs combine the power of generative adversarial networks (GANs) with the privacy-preserving technique of differential privacy (DP).
In this paper, we introduce an indoor localization framework employing DPGANs in order to generate privacy-preserving indoor location data.   
We evaluate the performance of our framework on a real-world indoor localization dataset and demonstrate its effectiveness in preserving privacy while maintaining the accuracy of the localization system.


\end{abstract}

\section{Introduction}  \label{sec:rw}
In recent years, IoT devices have become increasingly popular and have been developing quickly for smart cities, homes, health, agriculture, and factories~\cite{8762112}, \cite{privacyS22}. It has become extremely common for people to use indoor location-based services (LBSs) employing their smart devices~\cite{s10846}. Today, most people spend 80\% of their time in indoor areas while they actively use their smart devices. Three in four users of smart devices (74\%) utilize location-based applications~\cite{9531633}. 
To provide location-based services to users, such as guiding them through unfamiliar environments like hospitals, shopping malls, airports, and railway stations, indoor LBSs must rely on precise indoor localization. As a result, suitable indoor localization is essential for indoor LBSs to function effectively~\cite{s10846}. 

The most popular technique of indoor localization is location fingerprinting, which consists of two phases, training and localization~\cite{Access19}. In the training phase, the indoor localization system is trained with the fingerprints of various locations in the desired indoor area. The fingerprints here is referred to features of locations, such as received signal strength (RSS) from existing signal emitters. In the localization phase, the fingerprints of users are given to the formerly trained model as input so that it could monitor and localize them in that area.

Indoor localization systems can be location-based or zone-based. Location-based localization provides $(x, y)$ coordinates, while zone-based localization gives a zone. A zone in an indoor area is referred to a room, corridor, or any smaller part of the whole indoor environment.
Indoor localization systems often require accurate location data to provide useful services in various applications, such as navigation, asset tracking, security monitoring, and other location-based recommendations. However, the collection and use of this data raise various privacy and security concerns as it can reveal information about the user's activities, habits, and movements, which can be used to infer personal information about the user~\cite{FATHALIZADEH2022IEEE}. This information can be then used for targeted advertising or other purposes without the user's knowledge or consent. In indoor localization systems, user privacy can be violated in several ways:
(1) \textit{Location Tracking}: Indoor localization systems track the location of a user inside a building. This location data can be sensitive and can reveal a user's activities, habits, and interests. If this data falls into the wrong hands, it can be used to track and monitor users without their knowledge or consent~\cite {s10846}.
(2) \textit{Identification}: Indoor localization systems can also be used to identify individual users based on their unique location patterns~\cite{FATHALIZADEH2022102665}. For example, if a user's location data is collected over a period of time, their daily routine and habits can be identified. This can lead to targeted marketing, personal profiling, and even discrimination.
(3) \textit{ Re-identification}: Even if the user's identity is not initially known, it may be possible to re-identify the user by combining the location data with other information~\cite{FATHALIZADEH2022102665}. For example, if an attacker has access to social media data or other publicly available information, they may be able to link the location data with other data sources to re-identify the user.
(4) \textit{Inference}: In some cases, the location data can reveal sensitive information about the user even without revealing their exact location~\cite{JiangZG22}. For example, if a user frequently visits a specific medical facility, this may reveal sensitive information about their health condition.

There are ongoing efforts to develop techniques for protecting the privacy of individuals while still enabling the use of location-based services in indoor environments, such as anonymization, encryption, differential privacy (DP), and federated learning~\cite{FATHALIZADEH2022102665}, \cite{prihorus}, \cite{DPIndoor2022}, \cite{FATHALIZADEH2022IEEE}, \cite{9960135}. However, one of the most attractive solutions for protecting data privacy and mitigating possible attacks that have not been considered for indoor location privacy is to employ generated synthetic data instead of real data for training Artificial Intelligence (AI) models and analyzing data~\cite{10.14778/3231751.3231757}. This can be done by Generative adversarial networks (GANs)~\cite{goodfellow2014generative}. 
These networks can learn the training data distribution and generate synthetic data with a distribution very similar to the training data~\cite{Torkzadehmahani2019}. They are a way of creating a generative model through competition between two neural networks. The generator network takes in random noise as input and transforms it into synthetic data that resembles real data. The discriminator network then tries to determine whether the data it receives is real or generated by the generator. The generator network is trained through its competition with the discriminator to produce more and more convincing synthetic data~\cite{goodfellow2014generative}.
The GAN framework is a highly effective learning model that has been utilized for various applications, including but not limited to imitating expert policies and domain transfer~\cite{9158374}. 

Similar to other types of machine learning, GAN frameworks are also vulnerable to information leakage. In particular, the generator model tries to predict the underlying distribution of a dataset and produces realistic examples at random. This implies that the generator has the ability to recall training samples by means of deep neural networks. As a result, if the GAN model is employed on a private or confidential dataset, it may potentially expose the privacy of the dataset. To confront this issue, a potent solution, the GAN-obfuscator, is introduced in~\cite{Xu2019GANobfuscatorMI}. This framework employs a differentially private GAN (DPGAN) approach that introduces well-crafted Gaussian noise to the learning model gradients during the training phase. With the GAN-obfuscator in use, limitless synthetic data can be produced for any task, all while safeguarding the privacy of the training data~\cite{9158374}.

Despite the fact that private GANs are getting increasingly popular, their effectiveness on indoor location data remains unclear, which is the main focus of this work. The main contributions of this paper are listed as follows: 
\begin{itemize}
    \item To the best of our knowledge, this is the first work that introduces DPGANs for generating private indoor location data for both location-based and zone-based indoor localization.
    \item Our proposed DPGAN framework for indoor localization not only preserves the privacy of indoor location data but also enhances the accuracy of localization at the same time.
    \item We investigate the influence of two popular DPGANs for indoor location synthetic data, Differentially Private Wasserstein GAN (DPWGAN) and Differentially Private Conditional GAN (DPCGAN), and analyze the similarity of the generated datasets to the original dataset, the localization accuracy, and the privacy issues.
    \item The efficiency and the performance of the suggested DPGAN framework for indoor localization are verified via a real-world experimental testbed.
\end{itemize}

The rest of the paper is organized as follows. Section~\ref{sec:rw} discusses related studies on users' privacy preservation methods in indoor areas. In Section~\ref{sec:preliminary}, the preliminary concepts about indoor fingerprinting systems, DP, GANs and two DP-based GANs are explained. The suggested framework for implementing DPareGANs for indoor location data is presented in Section~\ref{sec:method}. The experimental setup, evaluation metrics, and model parameters is discussed in~\ref{sec:setup}. 
Section~\ref{sec:eval} is devoted to the performance evaluation of the proposed framework employing DPGANs for generating privacy-preserving indoor location data and Section~\ref{sec:conclusion} concludes the paper. 
\section{Related Work} \label{sec:rw}
It has been less than a decade that indoor location data privacy has been getting more attention. Authors in~\cite{holcersurvey}, summarize some of the studies on the privacy of users in indoor areas. A number of techniques have been developed to counter threats to indoor location privacy such as anonymization techniques, cryptography, differential privacy (DP) and local differential privacy (LDP), and federated learning (FL). Here, we briefly summarize existing studies using each of these techniques.

Among anonymization techniques, \textit{k}-anonymity is mostly employed for indoor location privacy protection. It hides the user's identity from the attackers in $k-1$ other users. Employing other anonymization approaches, \emph{e.g.}, $\ell$-diversity, $t$-closeness, $(\alpha,k)$-anonymity and $\delta$-presence, has been recently suggested in~\cite{FATHALIZADEH2022102665} to mitigate the attacks for \textit{k}-anonymity. 

Cryptography is also utilized in indoor localization~\cite{prihorus}, \cite{li_infocom}, in which where users apply homomorphic encryption on their RSS vectors and then send them to the server. Authors in~\cite{8806758}, on the other hand, deploy secure multi-party mechanisms and combine two secure two-party computing protocols to develop PPIL circuits construction components. The problems with these two techniques are heavy computation and communication overhead as users are mostly moving and their positions are continuously updated~\cite{9130897}. 

In addition to the aforementioned techniques, DP is being increasingly used to guarantee privacy regardless of adversary background knowledge~\cite{8493532}. Since DP is basically added to the RSS vectors of users, it leads to a considerable localization error. Recently, Adp-FSELM scheme is proposed in~\cite{DPIndoor2022} employing differentially private Wi-Fi and Bluetooth fingerprints fusion with semi-supervised extreme learning machine for indoor localization via a trusted aggregation in edge nodes, which also needs the participation of the high number of people. In order to mitigate the problems of DP, such as requiring a trusted server and a large number of data points, LDP is introduced for indoor positioning systems in~\cite{8736750}, \cite{LDPIndoor}, where no trusted server is necessary. Specifically, the combination of $k$-anonymity and LDP has suggested a paradigm-driven LDP framework for indoor localization in~\cite{8542955}. LDP in indoor positioning, nevertheless, requires a large number of participants, while there might not be that number of users in indoor areas.

Furthermore, authors in~\cite{FATHALIZADEH2022IEEE} propose indoor Geo-indistinguishability, which brings Geo-indistinguishability (that was originally introduced for outdoor positioning~\cite{geo}) into indoor localization. It is a DP-based privacy-preserving framework in a Geo-indistinguishability setting that protects the indoor location privacy of users by obfuscating the location data before leaving the users’ devices.

Finally, FL is also applied as a recently proposed privacy-preserving mechanism in indoor localization~\cite{9960135}, \cite{li2020pseudo}. Using FL provides privacy advantages because it is distributed and requires clients to train local models instead of transferring data to the central server~\cite{ciftler2020federated}. The authors in~\cite{liu2019floc} present FLOC utilizing a homomorphic encryption-based federated learning framework. Another work proposes personalized federated learning due to the fact that the fingerprint data is actually non-IID data as users usually move around in the indoor areas~\cite{wu2021personalized}. However, there is a high risk of getting users' information if attackers can access the weights of the FL-based system. Authors in~\cite{wu2021privacy} try to decrease this risk by applying DP on the model weights. In addition, employing FL is usually feasible when the indoor positioning system is trained based on the crowdsourcing technique, in which users are constantly involved in training to improve the localization model, which is not always the case. 

None of the aforementioned papers considers the applicability of DPGANs in indoor environments, in which there is no need for another trusted party and the presence of any other participants. Moreover, unlike other privacy-preserving mechanisms, DPGANs enable indoor positioning systems to be simultaneously private and accurate.

\section{Preliminaries} \label{sec:preliminary}
In this section, preliminary concepts used in this study including \emph{(i)} location fingerprinting in indoor environments, \emph{(ii)} DP, \emph{(iii)} GAN descriptions including WGAN and CGAN, and \emph{(iv)} two of commonly-used DPGANs (DPWGAN and DPCGAN) are explained.

\subsection{Fingerprinting Localization}
The location fingerprinting technique is inspired by the unique fingerprinting of people. Just as every individual has their own specific fingerprint, every location has also its own particular signal features. Depending on what kind of signal(s) we are receiving at each location, the signal features can be the RSS values. The most popular type of signal used in indoor localization is Wi-Fi signals.

\begin{figure*}[t]
    \centering
    \includegraphics[width=1\textwidth]{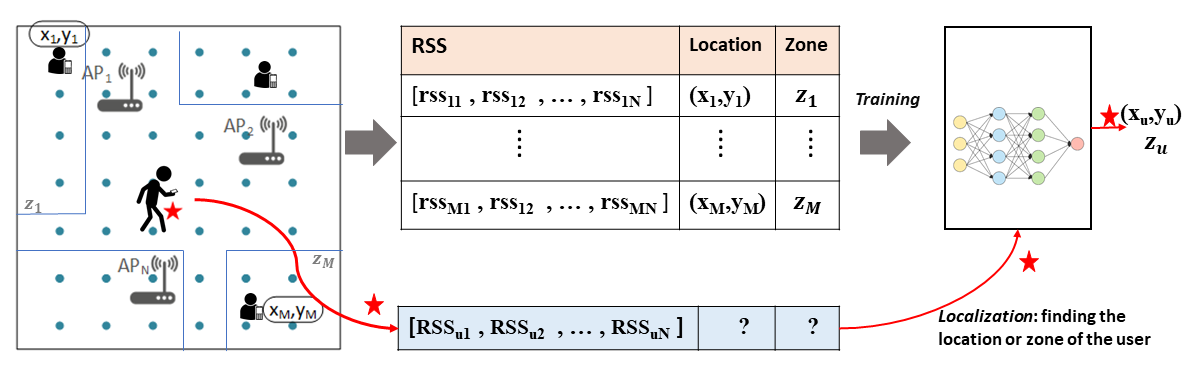} 
    \caption{The general overview of indoor location fingerprinting learning and localization processes.} 
    \label{fig:IPS}
\end{figure*}

Indoor fingerprinting is the process of creating a map of signal strengths or other features detected by sensors located at different points within an indoor environment. The purpose of this map is to estimate the location of a user's device based on the signals it receives from these sensors. Fingerprinting localization, in general, involves two stages: training and localization. Both of these phases make use of existing Wi-Fi Access Points (APs) as signal sources. Any device that is capable of detecting Wi-Fi signals can be utilized as a receiver~\cite{7131436}. 
The indoor positioning system is mainly trained by conducting surveys of the targeted area or getting huge amounts of data from people around that area to identify significant characteristics. The RSS values collected from existing APs at the known Reference Points (RPs) along with the $(x,y)$ coordinates of the RPs are stored in the radiomap on the server.

The recorded radiomap used for the AI training has a $M \times N$ dimension, where $M$ and $N$ are the numbers of RPs and APs, respectively. The fingerprint at a location $(x_i,y_i)$ is denoted by ${rss}_{i}^{RP} = [{rss}_{i1}, {rss}_{i2}, \ldots ,{rss}_{iN}]$, where ${rss}_{ij}$ is the RSS of ${AP}_j$ at the \textit{i}th RP. The AI model is exploited after the training. Then in the online phase, the location fingerprint of the user is injected as an input to the AI model so that the localization of the user can be carried out and LSP can estimate the position of the user, $(x,y)$. 
The user's fingerprint is obtained as the vector of RSS values, $S_u = [{RSS}_{u1}, {RSS}_{u2}, \ldots, {RSS}_{uN}]$, where ${RSS}_{uj}$ is the RSS of $AP_j$ received by the user. 
Machine learning algorithms for fingerprinting localization include traditional machine learning (ML), deep learning (DL), and deep reinforcement learning algorithms~\cite{MLIndoorSurvey}. Figure~\ref{fig:IPS} illustrates the indoor fingerprinting localization system with training/localization phases.

\subsection{Differential Privacy}
\textbf{Differential privacy (DP)}~\cite{10.1561/042} is commonly recognized as the go-to privacy analysis method. Depending on the degree of trust in the server, DP can be enforced either in a \textit{local} or \textit{global} context. By adding a suitable amount of noise to real-time or statistical data, DP ensures that it stays confidential while preserving an acceptable equilibrium between precision and privacy. In DP, if two datasets differ in at most one data point (one row), they are called \emph{adjacent}. Let privacy budget, {$\varepsilon \in \mathbb{R}_{\geq 0}$}, and $f$ be an algorithm operating on datasets. We say $f$ satisfies $\varepsilon$-Differential Privacy if for adjacent datasets $D, D'$ and all sets of possible outputs $M$, we have:
\begin{equation}
    \mathbb{P}(f(D) \in M) \leq \textrm{\emph{e}}^{\varepsilon}\mathbb{P}(f(D') \in M).
\end{equation}
By ensuring that the probability distributions on the output space originating from two input datasets cannot differ too much, $\varepsilon$-DP provides plausible deniability about any row's true value, even if all other rows are compromised. The lower $\varepsilon$, the stronger privacy $\varepsilon$-DP guarantees. To ensure privacy in the regressor/classifier learning setting, we demand that a regressor/classifier training algorithm satisfies $\varepsilon$-DP. 

\begin{figure*}[t]
    \centering
    \includegraphics[width=.8\textwidth]{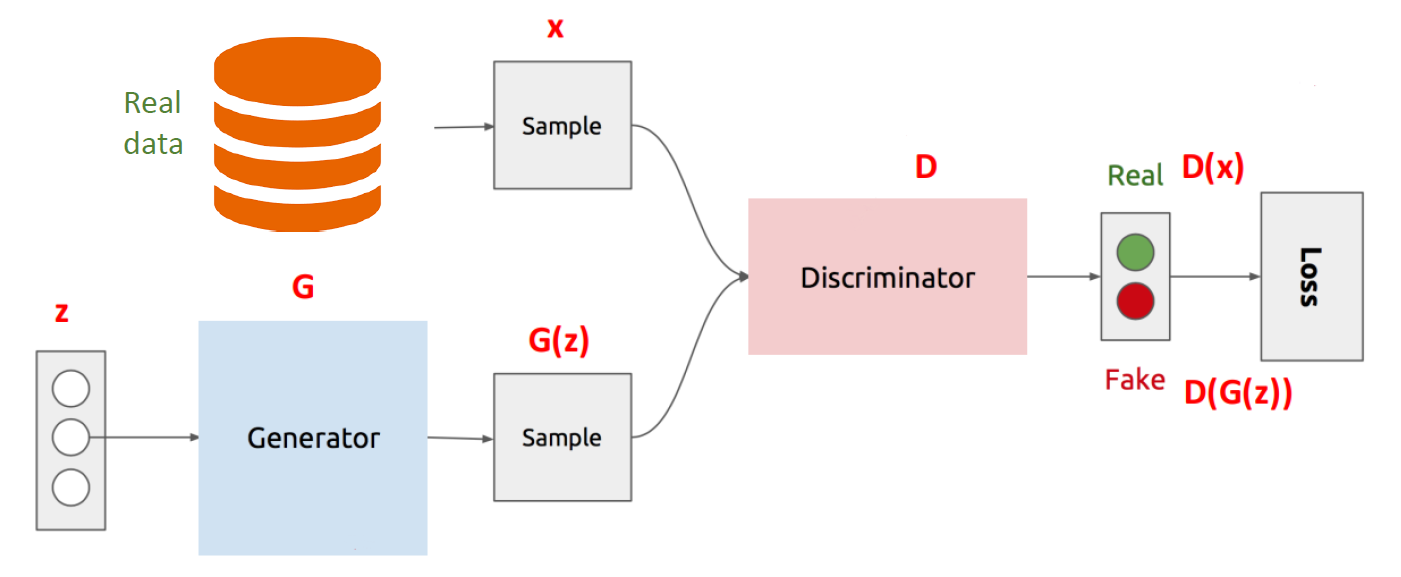} 
    \caption{The schematics of a GAN } 
    \label{fig:GAN}
\end{figure*}

\medskip
\textbf{$(\varepsilon, \delta)$-differential privacy~\cite{11761679_29}} A randomized mechanism ${M}$ is $(\varepsilon, \delta)$-differential private if for any set $S$ and any two neighbor databases $D$ and $D'$
\begin{align}
    & P( {M}(D) \in S ) \leq e^{\varepsilon}  P( {M}(D') \in S ) + \delta 
\end{align}
The Gaussian mechanism has been widely used to achieve $(\varepsilon, \delta)$-DP. In this model, for a function $f: D \rightarrow R$, its $L_2$ sensitivity is defined as $\Delta_2 f = max_{D, D'} \| f(D) - f(D') \|_2$. This function satisfies $(\varepsilon, \delta)$-DP, if Gaussian noise is added to $f$'s output, i.e., ${M}(D) = f(D) + \mathcal{N} (0, (\sigma \Delta_2 f)^2 )$ for $\varepsilon < 1$ and $\sigma > \sqrt{ 2 \ln 1.25 / \delta } / \varepsilon$.

\medskip 
\textbf{Rényi differential privacy~\cite{8049725}} is a privacy framework that quantifies the privacy guarantee provided by a randomized algorithm in terms of the divergence between the distribution of the output of the algorithm on neighboring inputs.
Formally, a randomized mechanism ${M}$ satisfies $\varepsilon$-Rényi differential privacy ($\varepsilon$-RDP) if for all pairs of neighboring datasets $D, D'$ and all $\alpha > 0$, the following inequality holds:
\begin{align}
    & \ln\left(\mathbb{E}[e^{\alpha M(D)}]\right) - \ln\left(\mathbb{E}[e^{\alpha M(D')}] \right) \leq \varepsilon \frac{\alpha^2}{2}
\end{align}
where $\mathbb{E}[e^{\alpha M(D)}]$ and $\mathbb{E}[e^{\alpha M(D')}]$ denote the expected value of the exponential mechanism applied to datasets $D$ and $D'$, respectively. This definition means that for any pair of neighboring datasets, the probability of any output from the mechanism $M$ on the first dataset being generated is at most $e^{\varepsilon\alpha^2/2}$ times greater than the probability of the same output being generated on the second dataset. 


\subsection{GANs}
Generative Adversarial Networks (GANs), introduced by Goodfellow et al. in 2014~\cite{goodfellow2014generative}, are employed to generate synthetic data with maximum similarity to a given dataset by learning its features' distribution over time using a Minmax game.
As it is shown in Figure~\ref{fig:GAN}, there are two neural network models in GANs: a \textit{Generator}, $G$, which contains a generative model for generating synthetic data and a \textit{Discriminator}, $D$, which is an adversarial model for detecting fake data. First, a Generator creates synthetic data using noise inputs drawn from a "latent space". A Discriminator, on the other hand, gets both original data and this Generator data as inputs and classifies the Generator data as fake or real by comparing them with the original data. This process results in a loss, which is used to optimize both Generator and Discriminator. Hence, there is quite a battle between Generator and Discriminator. The Generator tries to fool the Discriminator by learning to produce synthetic data that cannot be distinguishable from real ones, while the Discriminator tries to discriminate the authenticity of data by calculating the probability that input comes from real data instead of the Generator. The mathematical model on the Minmax game in GAN can be formulated as below to minimize the performance of $G$ while maximizing the performance of $D$:

\begin{equation}\label{GAN}
\min_{G}\max_{D}\mathbb{E}_{\textbf{x}\sim p_{\text{data}}(\textbf{x})}[\log{D(\textbf{x})}] +  \mathbb{E}_{\textbf{z}\sim p_{\text{z}}(\textbf{z})}[1 - \log{D(G(\textbf{z}))}]
\end{equation}
where $\textbf{x}$ and $\textbf{z}$ are sets of random samples from the data probability distribution $p_{\text{data}}(\textbf{x})$ and the noise probability distribution $p_\textbf{z}(\textbf{z})$ (the \textit{prior}), respectively. The first term indicates how well $D$ works on the real data and the second term determines how bad $D$ works on the generated data.


The Generator and Discriminator are trained using a loss function that measures the difference between the Generator's output and the true labels (real or fake) of the input data. Typically, the Generator and Discriminator use different loss functions. During the training process, the Generator and Discriminator are trained iteratively in a loop. In each iteration, the Generator generates synthetic data and the Discriminator classifies the input data as real or fake. The loss function is used to update the weights of the Generator and Discriminator, and the process is repeated for a fixed number of epochs. 
There are also different variations of GANs, such as Wasserstein GAN (WGAN)~\cite{arjovsky17a} and Conditional GAN (CGAN)~\cite{GoodfellowPMXWO20}. 


\medskip
\textbf{Wasserstein GAN (WGAN)} has proposed to improve the training stability by minimizing the distance between $p_z$ and $p_{data}$ through the following objective function:
\begin{align}
    & \min_{G}\max_{w \in W} \mathbb{E}_{x \sim p_{data}} [ f_w (x)] - \mathbb{E}_{z \sim p_z} [ f_w ( G(z))] 
\end{align}
where $\{ f_w \}_{w \in W}$ represent a family of K-Lipschitz functions for some constant $K$, that is $\| f \|_{Lip} \leq K$. 

\medskip
\textbf{Conditional GAN (CGAN)} allows both Generator and Discriminator to be conditional on the side of class labels, denoted by $y$, such that CGAN generates synthetic data for a given class. The objective of CGAN is computed as:
\begin{equation}
    \begin{aligned}
         \min_{G}\max_{D}  & \mathbb{E}_{\textbf{(x,y)}\sim p_{\text{xy}} }[\log{D(x, y)}] + \\
        & \mathbb{E}_{\textbf{z}\sim p_{\text{z}}  , \textbf{y}\sim p_{\text{y}}  } [ \log  (1- D(G(z, y) ) , y) ] \label{eq:cgan}
    \end{aligned}
\end{equation}
where $p_{xy}$ is the joint distribution from real samples $x$ and class labels $y$, and $p_y$ is the label distribution. 

\subsection{DPGANs System Model}
DPGANs can provide a way to balance privacy and accuracy by generating synthetic data that is statistically similar to the real data but does not reveal specific information about any specific data points. This can enable LBS to provide accurate services without compromising user privacy. Therefore, here we provide details on two re-known DPGANs, DPWGAN and DPCGAN, which are evaluated in this paper as the benchmark synthesizers for generating private indoor location data. 

Following work in~\cite{Abadi2016}, a number of studies utilized DP-SGD and GANs to generate differential private synthetic data~\cite{arxiv.1802.06739}, \cite{Torkzadehmahani2019}. In DPGANs, the DP mechanism is used to add noise to the gradients of the Generator during the training process to protect the privacy of the training data. The privacy budget in DP refers to the amount of noise that is added to the data, which is controlled by $\varepsilon$. In general, as the value of $\varepsilon$ decreases, the amount of noise added to the data increases, which can affect the quality of the generated data.

\subsubsection{DPWGAN}
The DPWGAN~\cite{arxiv.1802.06739} model enforces DP by incorporating noise into the Discriminator during training. Since differential privacy guarantees post-processing privacy, privatizing the Discriminator results in DP being enforced on the parameters of the Generator. This is because the mapping function between the Generator and Discriminator does not involve any private data. DPWGAN is based on WGAN and adds noise to the gradients while only clipping the model weights to maintain the network's Lipschitz property. These models have been tested on image data and electronic health records in previous studies.

Formally, a batch of prior samples $\{ z^{(i)} \}^m_{i=1} \sim p(z)$ and a batch of real data points $\{ z^{(i)} \}^m_{i=1} \sim p_{data} (x)$ are sampled. 
For each $i$, the gradients of the Discriminator are updated as follows:
\begin{align}
    & g_w (x^{(i)}, z^{(i)} ) \leftarrow \nabla_w [f_w(x^{(i)}) - f_w(g_{\theta} (z^{(i)})) ] 
\end{align}
the Gaussian noise is then added to the average of gradients as follows:
\begin{align}
    & \Bar{g_w} \leftarrow \frac{1}{m} (\sum^m_{i=1}   g_w (x^{(i)}, z^{(i)})  + \mathcal{N} (0, \sigma_n^2  c^2_g  I) )
\end{align}
and the weights on the next iteration are updated as below:
\begin{align}
    & w^{(t+ 1)} \leftarrow w^{(t)} + \alpha_d \cdot RMSProp (w^{(t)} ,  \Bar{g_w} ),\\
    &\quad w^{(t+ 1)} \leftarrow  clip (w^{(t+ 1)}, -c_p, c_p  ) 
\end{align}

The Generator then updates its gradients using the data received from a differentially private Discriminator. The network is as a whole differentially private due to the post-processing property of DP. 

\subsubsection{DPCGAN}
DPCGAN~\cite{Torkzadehmahani2019} has been proposed to generate synthetic data as well
as corresponding labels. In this setting, similar to DPWGAN, the differential private noise is added to the Discriminator's gradients based on the objective function of CGAN (Equation~\ref{eq:cgan}).  
DPCGAN has two innovations: 1) the training process splits the Discriminator loss between real data ($\log D(x,y)$) and generated data ($\log(1- D(G(z,y),y))$); 2) RDP accountant is proposed to be applied instead of Moment accountant. Formally, the gradients of the Discriminator are updated when the Gaussian noise is added to them:
\begin{align}
    &  \frac{1}{m}  (\sum^m_{i=1}   g_w (x^{(i)}, z^{(i)}) +  g_{w'} (x^{(i)}, z^{(i)}) + \mathcal{N} (0, \sigma_n^2  c^2_g  I) )
\end{align}
where $ g_w$ and $ g_{w'}$ are respectively the gradients of real and fake data, respectively.


\section{Indoor Location DPGAN framework} \label{sec:method}
This section introduces the proposed indoor location framework based on DPGANs. The motivation, the methodology, and the utility/privacy analysis of the proposed framework are discussed here.

\begin{figure*}[t]
    \centering
  \includegraphics[width=1\textwidth]{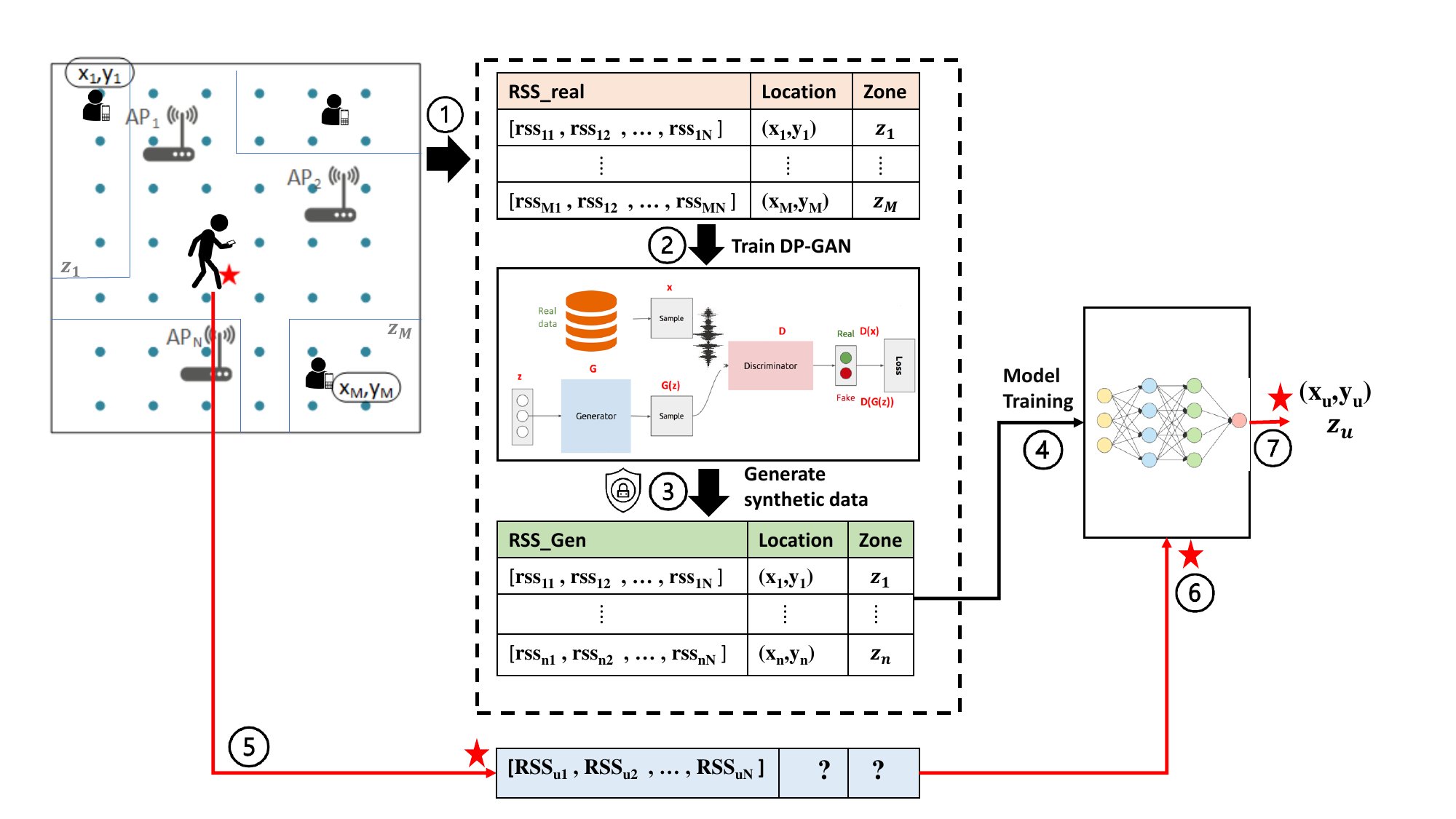}
    \caption{The proposed indoor location DPGAN framework.} 
    \label{fig:mainarc}
\end{figure*}

\subsection{Motivation}
The generated synthetic indoor location data can be used for a variety of tasks, such as testing and evaluating indoor positioning algorithms or simulating indoor environments for virtual reality applications. The generated data is differentially private, ensuring the privacy of individuals whose data was used to train the model, while still capturing the statistical patterns and relationships of the real-world data.

One possible motivation scenario for using DPGANs for indoor localization could be in a commercial shopping mall setting. Consider a shopping mall that intends to provide personalized recommendations to shoppers based on their current location and browsing history, but also needs to protect the shoppers' privacy. The mall has a network of beacons installed throughout the mall to track the location of shoppers, and collects this location data to build models for personalized recommendations. However, collecting the location data can be sensitive and can reveal information about the shoppers' activities and interests. To protect the shoppers' privacy while still providing accurate recommendations, the mall could use a variation of DPGANs to generate synthetic location data. The DPGAN could be trained on the real location data collected by the beacons and could generate synthetic data that is statistically similar to the real data but does not reveal information about specific individuals. This synthetic data could be then used to build the recommendation models without compromising the shoppers' privacy. DPGANs enable the shopping mall to provide personalized recommendations to shoppers while protecting their privacy, which could lead to increased customer satisfaction and loyalty.

\subsection{Methodology}
The architecture of our indoor location DPGAN framework has been depicted in Figure~\ref{fig:mainarc} and it can be described in the following steps:

\textbf{Step \textcircled{1}}: The real location fingerprinting data is measured and gathered from people within a target indoor area with a crowdsourcing technique in order to create a dataset for that environment. For the location-based approach, the dataset contains RSS values and $(x,y)$ coordinates, while for the zone-based approach, the dataset consists of RSS values and zones.

\textbf{Step \textcircled{2}}: The dataset collected in the previous step is used to train the DPGAN. This dataset is considered input data to the Discriminator network to build a DL model for DPGAN. The training process of DPGAN is dependent on whether the indoor localization system is location-based or zone-based. For location-based localization, DPWGAN is used and trained on RSS values and $(x,y)$ location coordinates, while for zone-based localization both DPWGAN and DPCGAN can be employed and trained on RSS values and zones.

\textbf{Step \textcircled{3}}: The trained location-based or zone-based DPGAN is employed for generating synthetic data similar to the original data. It can also generate more synthetic fingerprinting data in order to improve localization accuracy and preserve privacy at the same time. 

\textbf{Step \textcircled{4}}: The synthetic dataset is used to train an ML/DL model for localization so that we could further localize new users in that environment. Based on the location-based or zone-based approaches of the indoor localization system, this model can be either a regressor or a classifier and the position of a new user, hence, is provided based on its $(x,y)$ coordinates or its zone.

\textbf{Step \textcircled{5}}: A new user arrives, who requires to receive services based on her/his position. The fingerprints of her/his place are then collected by her/his own devices.

\textbf{Step \textcircled{6}}: The RSS fingerprints of a new user are given to the trained localization model.

\textbf{Step \textcircled{7}}: Resting on whether the indoor localization system is location-based or zone-based, the location coordinates $(x_u,y_u)$ or zone $(z_u)$ of a new user is computed.

\subsection{Utility/Privacy Analysis}

Basically, DPGANs achieve their privacy by adding noise to the gradient updates of the Generator and Discriminator during the training process. 
This noise then helps to ensure that the output of the Generator is not dependent on any single input data point, which makes it more difficult for an attacker to infer any information about an individual's data from the generated dataset. 
Moreover, DPGANs provide formal privacy guarantees in the form of DP, \emph{i.e.}, they are designed to ensure that the probability distribution of the generated data does not change significantly when a single individual's data is removed from the original dataset. This means that an attacker would not be able to determine whether or not a particular individual's data was included in the original dataset based on the generated data alone.
On the other hand, the Generator network of DPGAN is designed to learn the underlying distribution of the original data and to produce synthetic data that is similar to it. This means that the network generates data with a distribution very similar to that of original data, which results in obtaining accurate outcomes for location-based services purposes. 
 
\section{Experimental setup} \label{sec:setup}
This section explains the real-world dataset employed to test our framework, the evaluation metrics, and the hyper-parameters of the ML models utilized in the proposed framework.

\subsection{Dataset}
\begin{figure}[t]
    \centering
    \includegraphics[width=.45\textwidth]{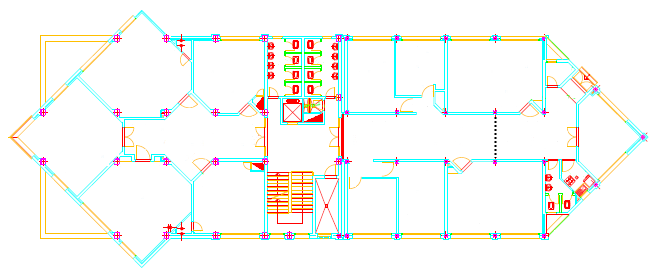} 
    \caption{The floor plan of the CRI testbed. } 
    \label{fig:Paj2D}
\end{figure}

Here, we employed CRI dataset which is measured in a $51m \times 18m$ testbed situated on the second floor of the Cyberspace Research Institute (CRI) at Shahid Beheshti University. The layout can be seen in Figure~\ref{fig:Paj2D}. There are nine Wi-Fi APs installed on the ceiling. The radio signal strength values of $384$ reference points were recorded from all nine access points in four directions, with $100$ samples taken in each direction, and saved in the radiomap database. Additionally, within a short period of time, $100$ inquiry samples were gathered at different times for the purpose of evaluating the experiment All $400$ samples at each point are averaged to get a more accurate fingerprint of them. The whole environment also consists of $15$ zones including rooms and corridors. Therefore, the dataset has $12$ columns including, $9$ RSS values from all APs, the $(x,y)$ coordinates of the points, and the zone number where the point is located. The details on this dataset can be found in~\cite{Access19}.

\subsection{Evaluation Metrics}
Here the performance of location-based and zone-based DPWGAN and zone-based DPCGAN in indoor locations are evaluated and compared with the performance of non-private WGAN and CGAN from two perspectives: data utility and privacy evaluation. 

In the utility assessment, generally, if synthetic data exhibit similar statistical characteristics and model accuracy as the original data, it is considered to have high utility. Therefore, the correlation between features in the original dataset is compared with feature correlations in the synthetics datasets. 
Furthermore, we employed various machine learning models to explore the usefulness of generated datasets in AI training activities in both location-based and zone-based modes measuring Root Mean Square Error (RMSE) in meters and accuracy (\%), respectively. 

In the privacy assessment, we assessed the effectiveness of our framework in preventing the disclosure attack. DPWGAN/DPCGAN is used to generate synthetic data based on the original data used to train them. Although it is expected that the records of real data will not appear in synthetic data, still we should be cautious about the possibility of accidental privacy leaks. This leakage might arise due to the unintentional memorization of some original records during training. To examine this matter, we can check for identical records in the generated datasets using a straightforward but practical approach based on Euclidean distance. To assess the potential memorization issue (disclosure risk), we measure the average Euclidean distance between the generated records and the closest records in the original dataset. This metric helps us determine how far a generated point is away from the original data points and whether this could result in privacy leakage.

\subsection{Model Structure and Parameters}
We have implemented all the experiments to evaluate our framework in Python. The experiments were performed on a machine with an Intel Corei$7$ $1.8$GHz CPU and $12$GB of RAM. 
In order to assess the datasets' quality, we use scikit-learn tools and employed a group of four models (MLP, Random forest, SVM, and Decision tree). These models are trained on the datasets and their predictive performance and reliability are evaluated by testing their ability to accurately predict the labels. The learning rate here is $10^{-4}$. Other parameters are the default parameters of the models in scikit-learn.

The training process also is based on the five-fold cross-validation. In other words, training the models involves dividing the generated dataset into five subsets, and then using each subset in turn for validation while training the model on the remaining four subsets. After training, the performance of the model is evaluated by testing it on the portion of the generated dataset, when having been also trained on a portion of the generated dataset.


\section{Empirical Evaluation} \label{sec:eval}
In this section, we assess our indoor localization DPGAN framework from data utility and privacy points of view. 

\begin{figure*}[!t]
	\centering
         \begin{subfigure}{0.31\textwidth}
\includegraphics[width=0.95\textwidth]{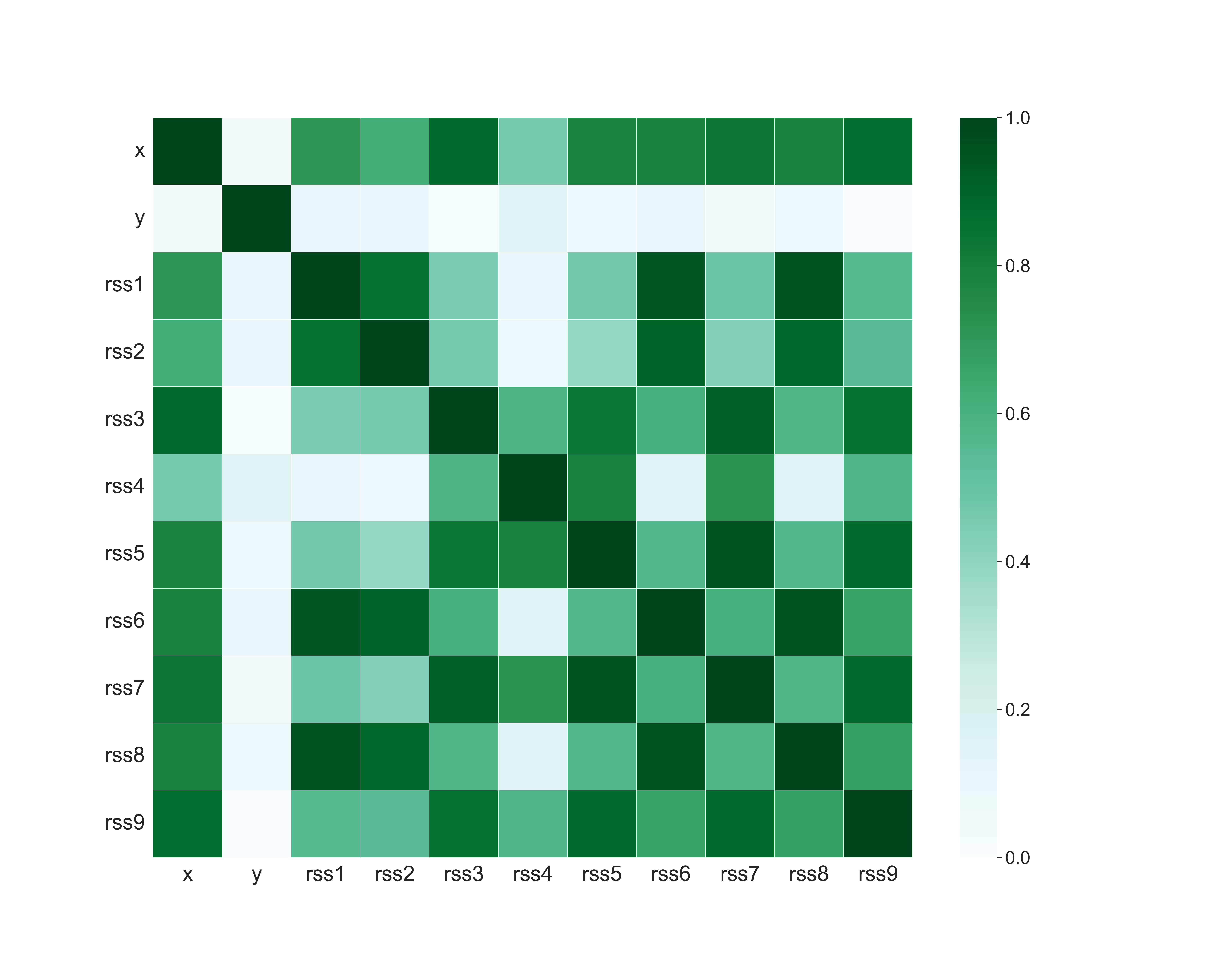} 
	\caption{Original}
 \vspace{0.25cm}
	\label{fig:Corr_original}
	\end{subfigure}
	\begin{subfigure}{0.31\textwidth}
	\includegraphics[width=0.95\textwidth]{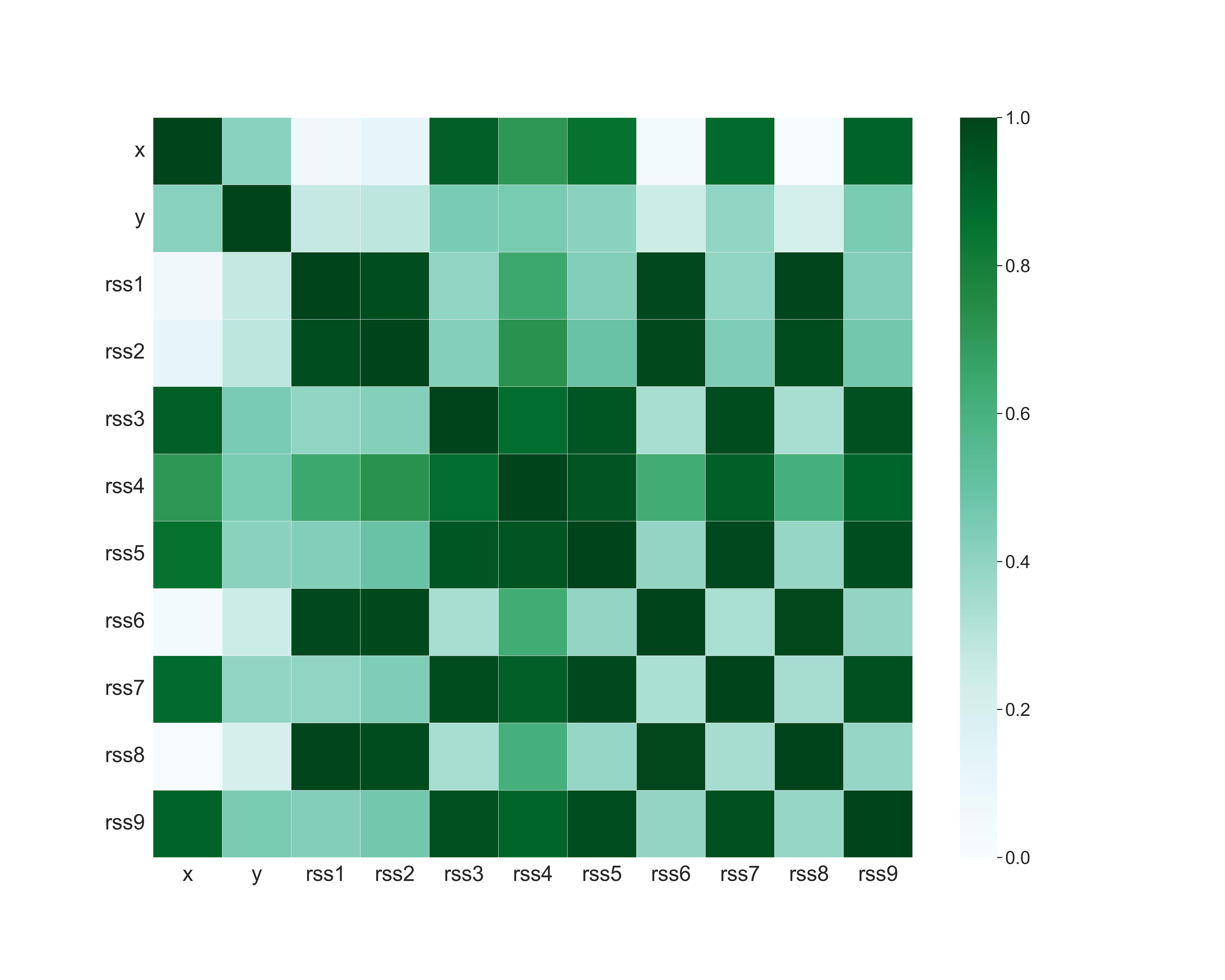} 
	\caption{Non-private WGAN}
  \vspace{0.25cm}
	\label{fig:Corr_wgan}
	\end{subfigure}
	\begin{subfigure}{0.31\textwidth}
	\includegraphics[width=0.95\textwidth]{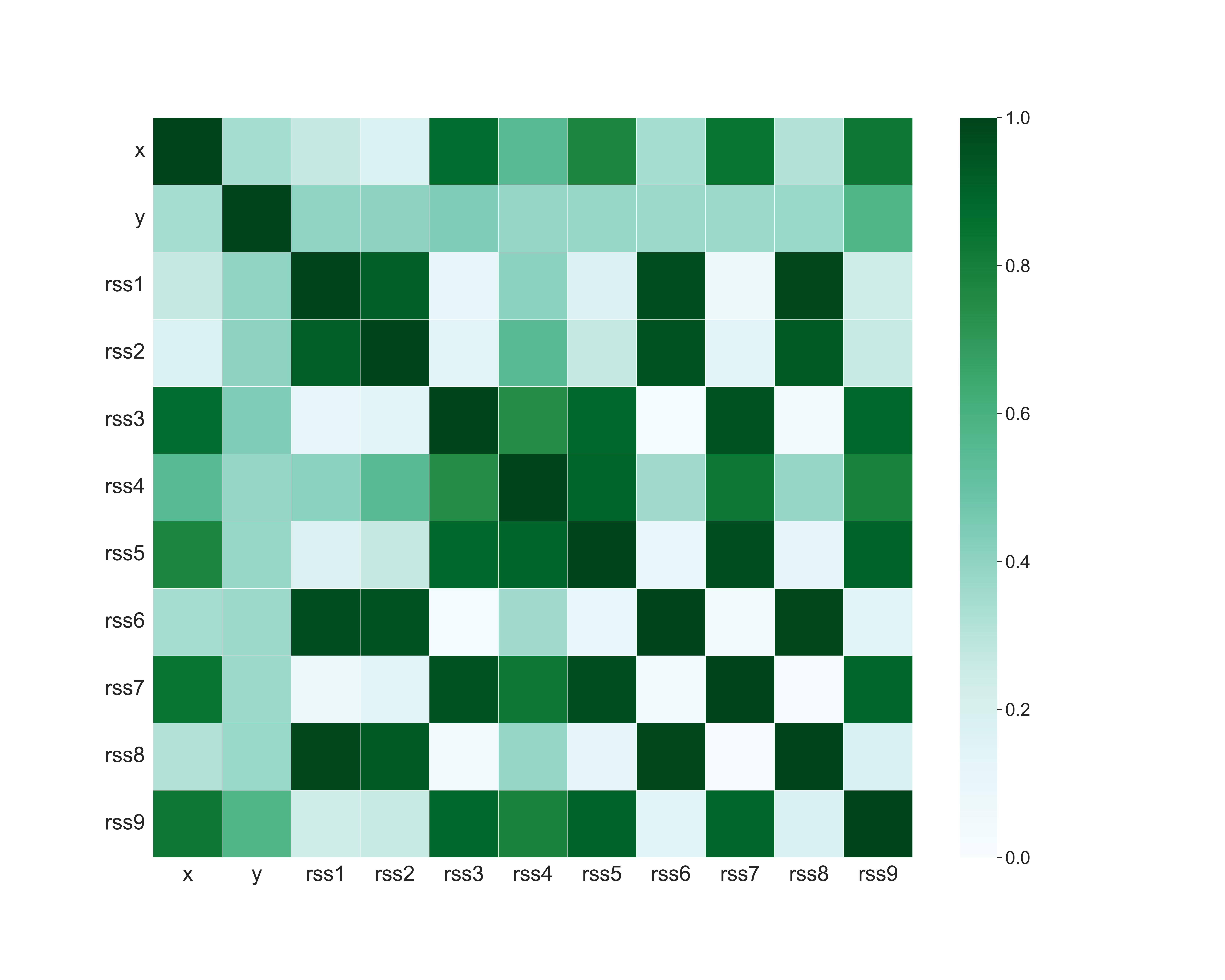}
	\caption{DPWGAN ($\varepsilon = 1$)}  
   \vspace{0.25cm}
	\label{fig:Corr_dpwgan_eps1}
	\end{subfigure}
	\caption{Correlation results for location-based DPWGAN} 
	\label{fig:Corr_dpwgan_reg}
\end{figure*}

\begin{figure*}[!t]
	\centering
         \begin{subfigure}{0.31\textwidth}
 \includegraphics[width=0.95\textwidth]{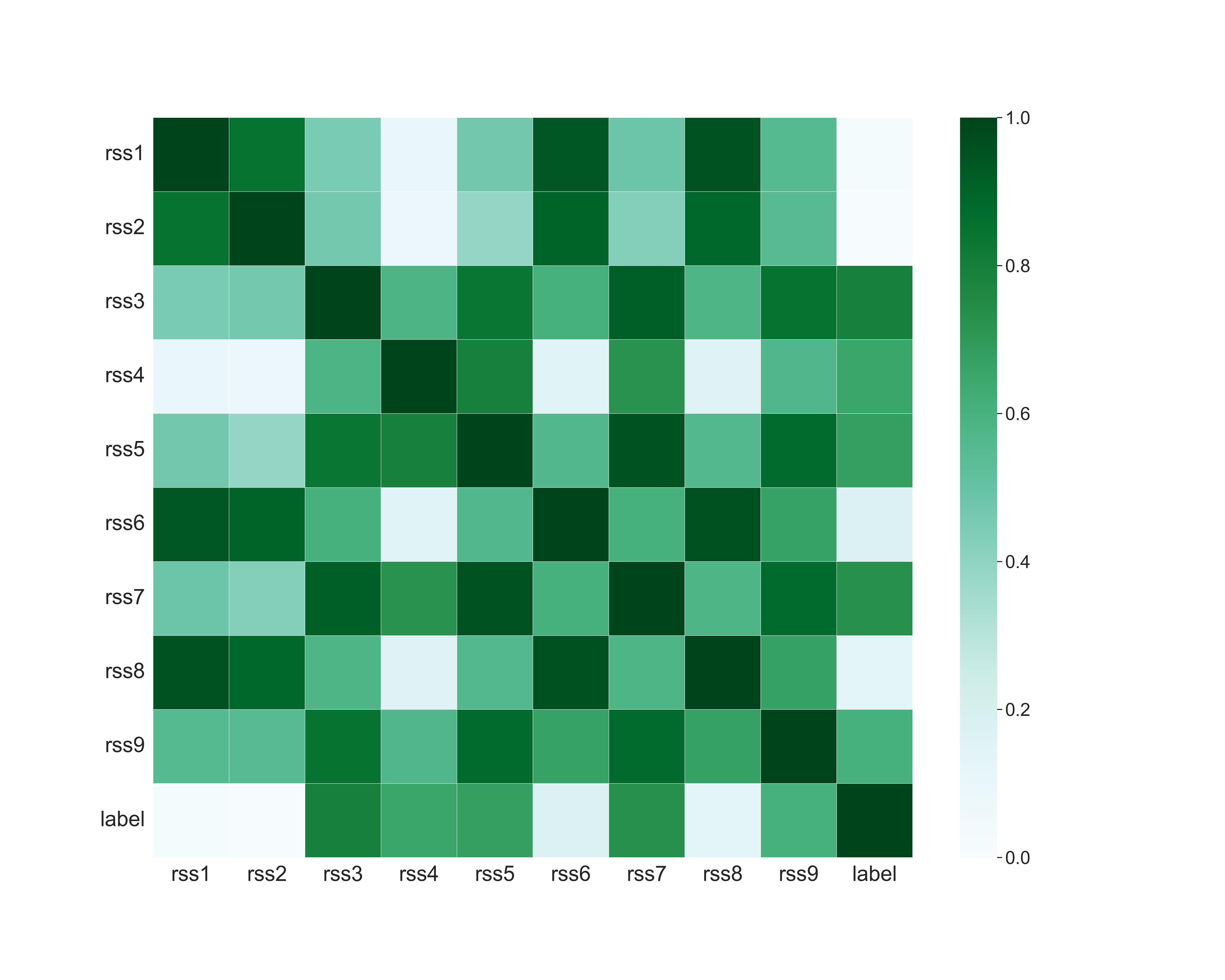} 
	\caption{Original}
 \vspace{0.25cm}
	\label{fig:Corr_original_dpwgan_class}
	\end{subfigure}
	\begin{subfigure}{0.31\textwidth}
 \includegraphics[width=0.95\textwidth]{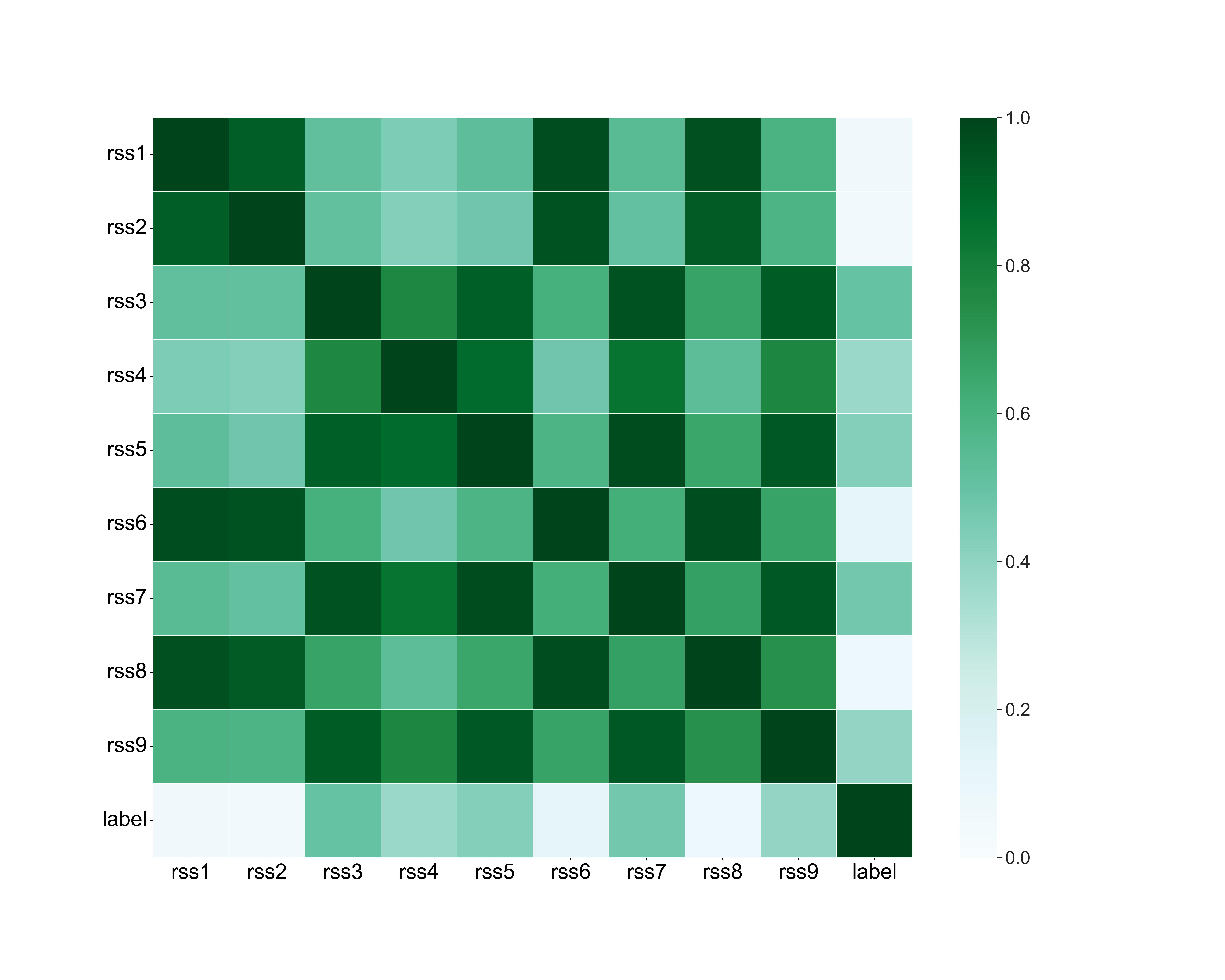} 
	\caption{Non-private WGAN}
  \vspace{0.25cm}
	\label{fig:Corr_wgan_class}
	\end{subfigure}
	\begin{subfigure}{0.31\textwidth}
	\includegraphics[width=0.95\textwidth]{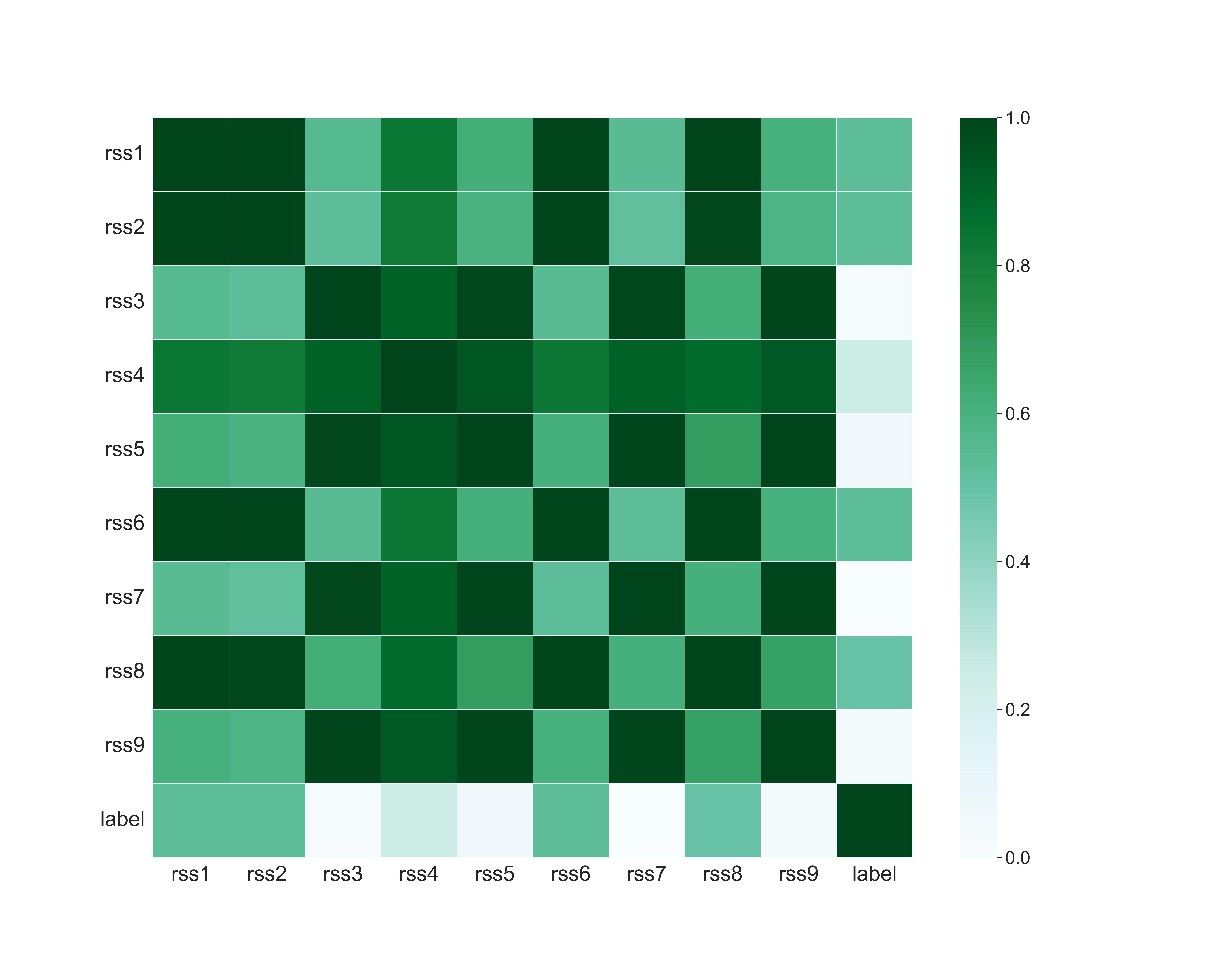}
	\caption{DPWGAN ($\varepsilon = 1$)}    	
   \vspace{0.25cm}
	\label{fig:Corr_dpwgan_eps1_class}
	\end{subfigure}
	\caption{Correlation results for zone-based DPWGAN  } 
	\label{fig:Corr_dpwgan_class}
\end{figure*}
\begin{figure*}[!t]
	\centering
         \begin{subfigure}{0.31\textwidth}
\includegraphics[width=0.95\textwidth]{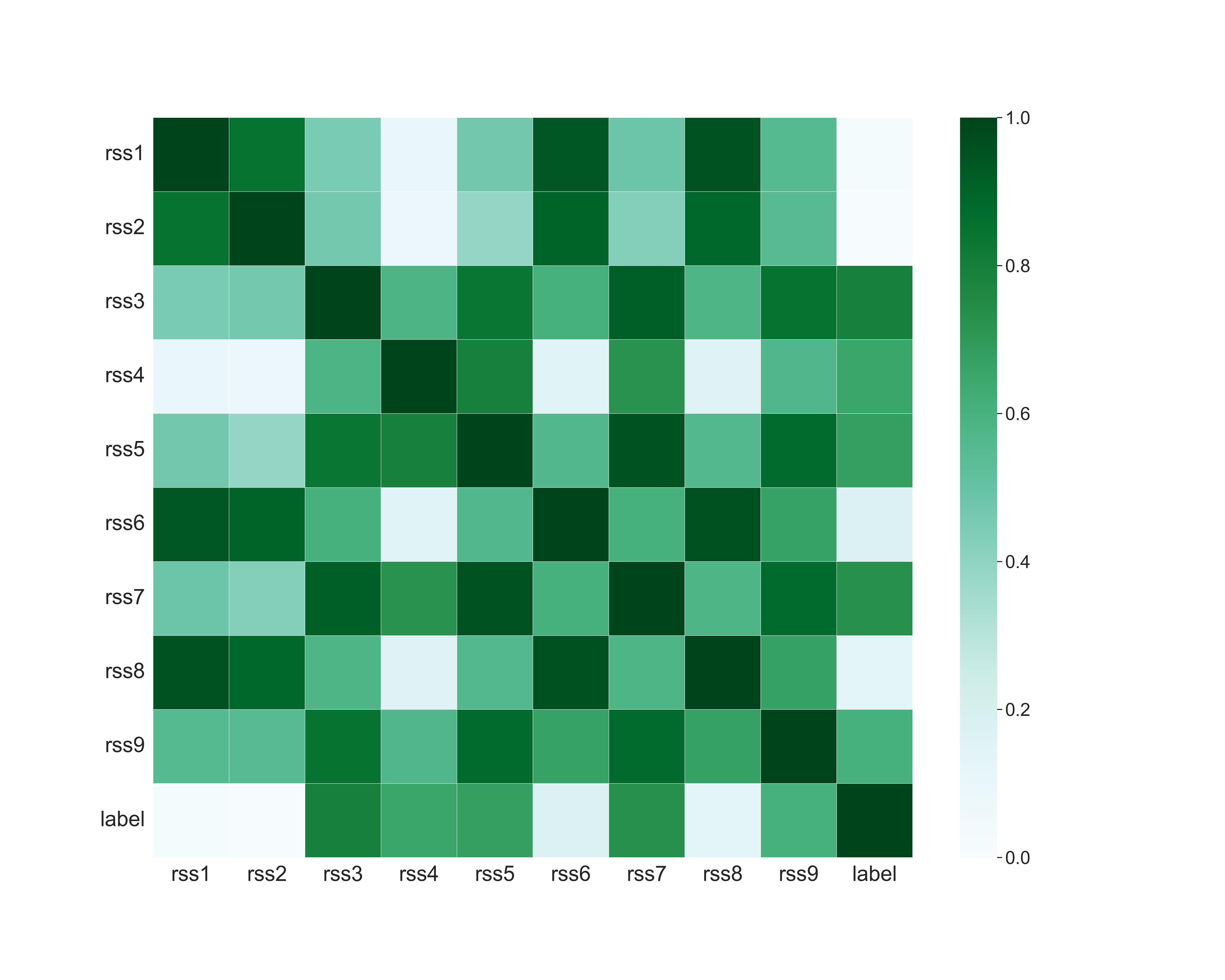}
	\caption{Original}
 \vspace{0.25cm}
	\label{fig:Corr_original_dpcgan}
	\end{subfigure}
	\begin{subfigure}{0.31\textwidth}
	\includegraphics[width=0.95\textwidth]{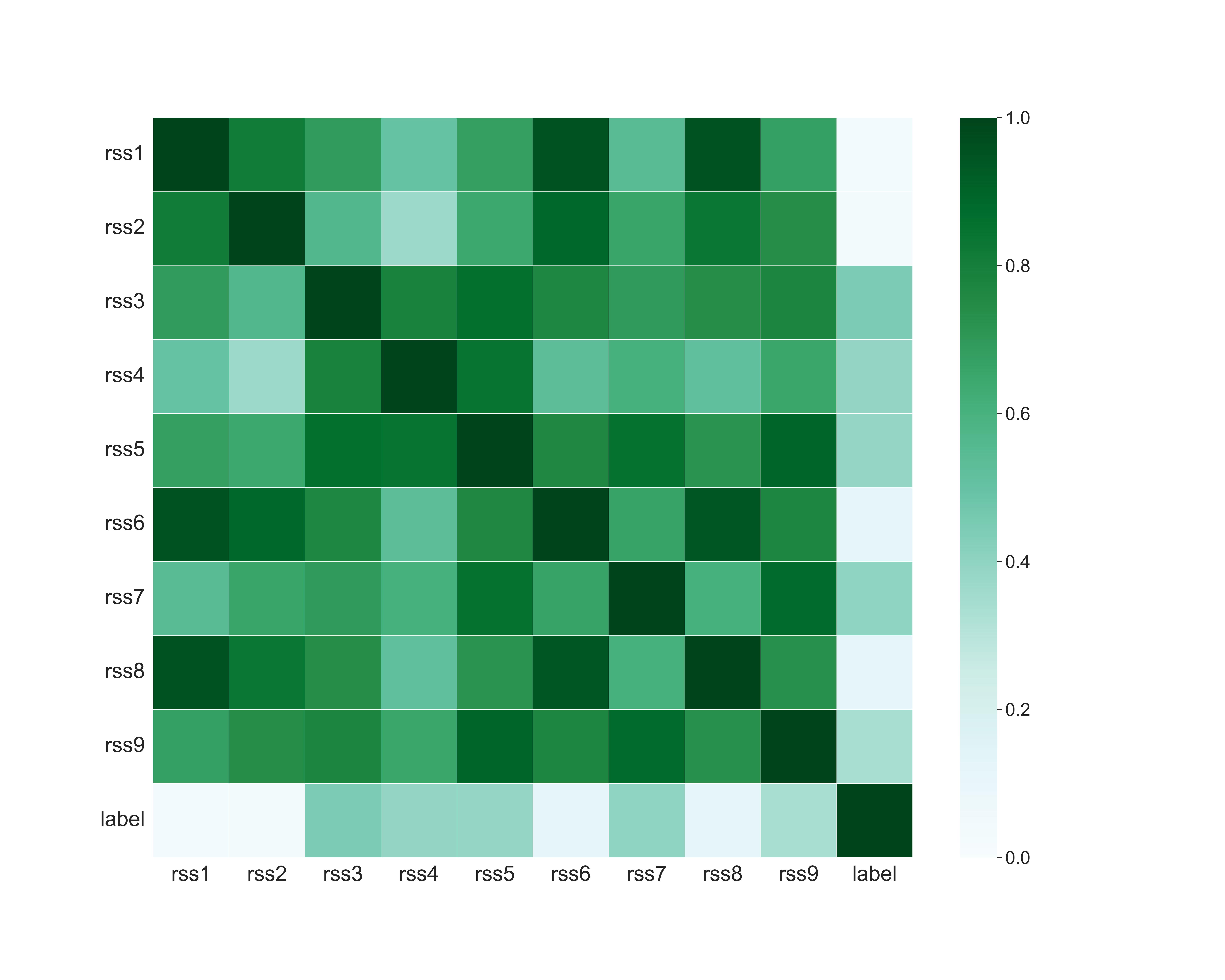} 
	\caption{Non-private CGAN}
  \vspace{0.25cm}
	\label{fig:Corr_cgan}
	\end{subfigure}
	\begin{subfigure}{0.31\textwidth}
	\includegraphics[width=0.95\textwidth]{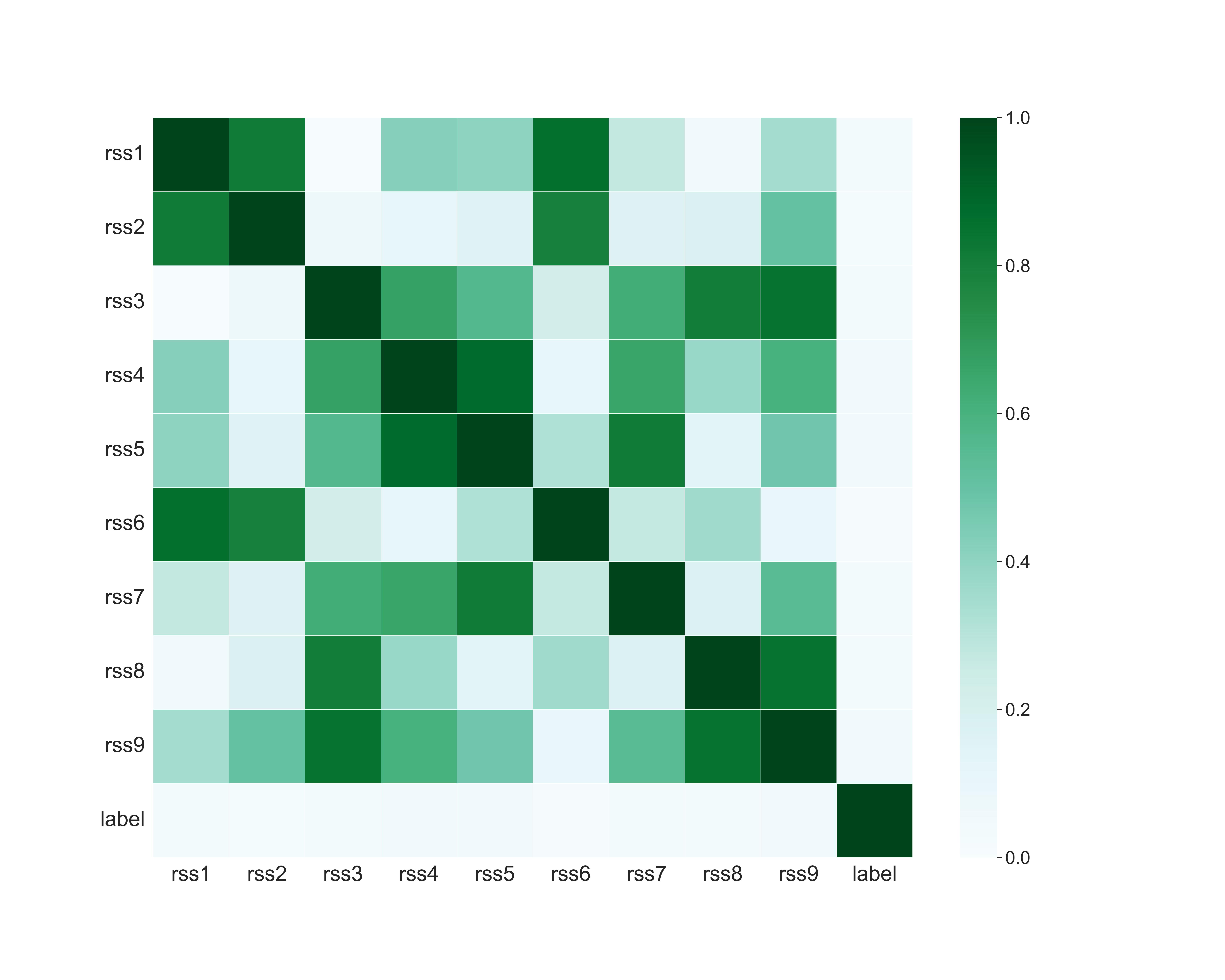}
	\caption{DPCGAN ($\varepsilon = 1$)}   
   \vspace{0.25cm}
	\label{fig:Corr_dpcgan_eps5}
	\end{subfigure}
	\caption{Correlation results for zone-based DPCGAN  } 
	\label{fig:Corr_dpcgan}
\end{figure*}

\subsection{Utility Evaluation}
The objective of the utility evaluation is to examine if the synthetic data produced by our framework maintains the correlations of the original data. To measure how the correlation between features in the original dataset is preserved compared to generated datasets, we compute the correlation matrix in terms of \emph{Pearson correlation coefficient}. This metric quantifies the strength and direction of the linear association between two variables. Its value ranges from $-1$ to $+1$, with $0$ indicating no correlation. A correlation coefficient of $1$ indicates a perfect positive linear relationship between the two variables, while a correlation coefficient of $-1$ indicates a perfect negative linear relationship. The closer the correlation coefficient is to $0$, the weaker the linear relationship between the variables.
Formally, the Pearson correlation coefficient between two random variables $X$ and $Y$ with means $\mu_X$ and $\mu_Y$ and standard deviations $\sigma_X$ and $\sigma_Y$, respectively, is defined as:
\begin{align}
    & \rho_{X,Y} = \frac{\operatorname{cov}(X,Y)}{\sigma_X \sigma_Y} = \frac{E[(X-\mu_X)(Y-\mu_Y)]}{\sigma_X \sigma_Y} ]
\end{align}
where $\operatorname{cov}(X,Y)$ is the covariance of X and Y, E is the expected value operator, and $\mu_X$, $\mu_Y$, $\sigma_X$, and $\sigma_Y$ are the means and standard deviations of $X$ and $Y$, respectively. In this study, we just consider the absolute values of correlation coefficients as the direction of correlation has no impact on our analysis. Thus, the coefficient values are reported in the range $[0,1]$.

Here, the Pearson correlation coefficient is used to measure the mutual correlation between each pair of features in \emph{(i)} the original dataset, \emph{(ii)} datasets generated using WGAN and CGAN, and \emph{(iii)} datasets generated using DPWGAN and DPCGAN ($\varepsilon = 1$). For DPWGAN, we have tried both location-based and zone-based localization, but DPCGAN is used only for zone-based positioning.
Figures~\ref{fig:Corr_dpwgan_reg}, \ref{fig:Corr_dpwgan_class}, and \ref{fig:Corr_dpcgan} depict the feature correlation results for location-based DPWGAN, zone-based DPWGAN, and zone-based DPCGAN, respectively. The darker color shows that there is a stronger correlation between the associated features (higher coefficient value). 

It can be observed that, in general, the trend of correlation remains consistent across each set of experiments. This means that if a pair of features have a strong correlation in the original dataset (indicated by darker colors), that pair of features have still strong correlation in datasets generated by non-private WGAN/CGAN and DPWGAN/DPCGAN ($\varepsilon = 1$). This trend, however, has been better preserved when location-based WGAN/DPWGAN has been used for generating synthetic data. This result suggests applying DPWGAN for generating location-based indoor location data when protecting the correlation between features is of crucial importance.  

\subsection{AI Training Performance}
Similar to what is done in~\cite{JiangZG22}, in order to assess the usefulness of synthetic data in various AI training endeavors, a variety of ML algorithms are employed. Therefore, we use the original dataset in comparison with the generated dataset by non-private GANs and DPGANs to train the ML model for step $4$ of our framework. For a high utility private data generation, it is expected to achieve similar accuracy to the accuracy obtained with the real dataset while it is private at the same time. Using indoor location datasets, we can train both regressors and classifiers to evaluate the performance of the private synthetic dataset on them.

\subsubsection{Regression results}
In the case of regression training, we remove the zone number column from the datasets and report the locations in $(x,y)$ coordinates. To evaluate the utility of our framework, we trained four regressors, namely MLP, Random Forest, SVM, and Decision Tree, on the original CRI dataset, and on datasets generated using non-private WGAN, and DPWGAN ($\varepsilon= 1, 5, 10, 15$) in a location-based mode. 
The utility has been measured based on the average distance in terms of RMSE (meters) between the real locations and those estimated by regressors trained on different synthetic data.  
Table~\ref{tab:dpwgan_reg} reports the RMSE results for the location-based localization. As it is expected, it can be seen that the utility of generated data reduces (RMSE increases), when more privacy constraint is required (lower $\varepsilon$ value). This trend is independent of the selection of regressor.  
It also indicates that the RMSE obtained by applying MLP and SVM regressors on the generated dataset by DPWGAN with $\varepsilon$ greater than $10$ are very similar to the RMSE by the original dataset and the non-private WGAN.

\begin{figure}[!t]
    \centering
    \includegraphics[width=.49\textwidth]{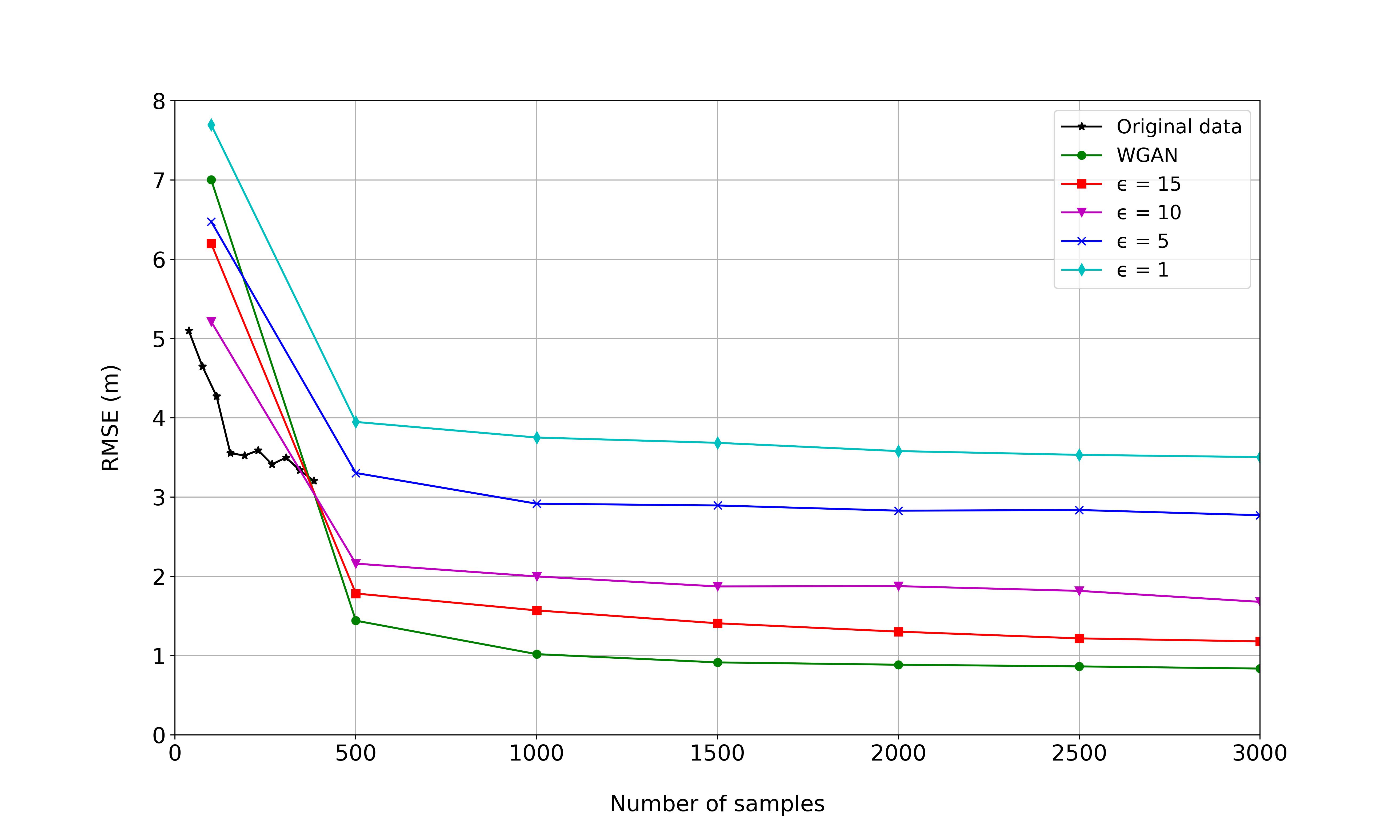}
    \caption{The location-based RMSE of MLP regressor using original dataset vs generated datasets by non-private WGAN and DPWGAN with $\varepsilon = 1, 5, 10, 15$ and $\delta = 10^{-5}$, based on the number of fingerprint samples.} 
    \label{fig:dpwgan_samples}
\end{figure}

Figure~\ref{fig:dpwgan_samples} depicts the RMSE of MLP regressor when it is trained on the original dataset (dashed black line) vs generated datasets by non-private WGAN and DPWGAN for $\varepsilon =1, 5, 10, 15$ and $\delta = 10^{-5}$ based on the number of fingerprints in the dataset.
The number of records in the original dataset is $384$, while for the synthetic datasets, we tried to generate more fingerprint records so that we could improve the localization accuracy as well. 
The results show that the accuracy of the regressor has been improved when it has learned the pattern from a larger dataset. This is specifically considerable up to the point that the number of records reaches 1000. For a higher number of records, the accuracy of the indoor location fingerprinting system goes to the saturated phase which is compatible with~\cite{Access19}.
Therefore, using DPWGAN not only provides a more private dataset but also decreases the localization error as more data could be generated for training. Even with more private synthetic datasets ($\varepsilon =$  1, 5), we can achieve the same or less localization error as the original dataset in a higher number of fingerprint samples.

\begin{table*}	[!t]
    \caption{The location-based RMSE of four regressors when using original dataset vs generated dataset by non-private WGAN and DPWGAN on $\varepsilon =1, 5, 10, 15$ and $\delta = 10^{-5}$ (for 384 records).  }%
    \centering
		\resizebox{0.85\textwidth}{!}{%
			\begin{tabular}{@{ }l r r r r @{ }}
				\toprule
                & \multicolumn{4}{c}{RMSE (m)} \\ \cline{2-5}  
                \multicolumn{1}{l}{} &
				\multicolumn{1}{c}{\textbf{MLP} } &
				\multicolumn{1}{c}{\textbf{Random Forest} } &
				\multicolumn{1}{c}{\textbf{SVM}} &
				\multicolumn{1}{c}{\textbf{Decision Tree}} 
				\\
				\midrule
				\textbf{Original} & 3.28	$\pm$ 0.2  & 2.36 $\pm$ 0.4 & 3.17 $\pm$ 0.2  & 3.29 $\pm$ 0.8  \\	
				\textbf{WGAN (non-private)} & 3.51 $\pm$ 0.2  & 2.46 $\pm$ 0.2 & 2.29 $\pm$ 0.3 & 4.30 $\pm$ 0.2 \\
                \textbf{DPWGAN ($\varepsilon = 15$)} & 3.38 $\pm$  0.2 & 3.75 $\pm$ 0.3 & 2.66 $\pm$  0.4 & 5.36 $\pm$ 0.2  \\
                \textbf{DPWGAN ($\varepsilon = 10$)} & 3.46 $\pm$ 0.2 & 3.98 $\pm$ 0.1 & 3.79  $\pm$ 0.2 & 5.59 $\pm$ 0.1  \\
				\textbf{DPWGAN ($\varepsilon = 5$)} & 4.22 $\pm$ 0.1  & 4.33 $\pm$ 0.1 & 3.81 $\pm$ 0.2 & 4.63 $\pm$ 0.1 \\
                \textbf{DPWGAN ($\varepsilon = 1$)} & 4.97 $\pm$ 0.2 & 4.63 $\pm$ 0.1 & 5.02 $\pm$ 0.4 & 5.58 $\pm$ 0.3 \\	
				\bottomrule
			\end{tabular}
		}
	
	\label{tab:dpwgan_reg}
\end{table*}

\subsubsection{Classification results}
Here, classification models are applied for zone-based indoor localization. In this case, zone numbers are considered as labels for the ML models and the locations are reported in zone numbers, not in $(x,y)$ coordinates. For classification purposes, thus, we remove $(x,y)$ coordinates from the dataset and generate the RSS columns and labels. In DPWGAN, this is done by considering the label as another feature and generating all features (RSS and label) simultaneously. In DPCGAN, which is conditioned by labels, the model is trained on the data associated with each label separately.

We again consider four classifiers including MLP, Random Forest, SVM, and Decision Tree and apply the original CRI dataset, and datasets generated using non-private WGAN, DPWGAN ($\varepsilon= 1, 5, 10, 15$), non-private CGAN, and DPCGAN ($\varepsilon= 1, 5, 10, 15$) on them, all in zone-based localization mode. Table~\ref{tab:dpwgan_class} and ~\ref{tab:dpcgan_class} list the zone-based accuracy of these four classifiers for DPWGAN and DPCGAN, respectively. As expected, in lower $\varepsilon$ values when privacy constraints are higher, the utility of the generated data reduces and we obtain lower accuracy. This trend is consistent with all classifiers.  
Table~\ref{tab:dpwgan_class} also specifies that the accuracies of MLP and SVM classifiers for the generated dataset by DPWGAN with $\varepsilon$ greater than $10$ are very similar to the accuracy by the original dataset and the non-private WGAN. Table~\ref{tab:dpcgan_class}, on the other hand, indicates the lower accuracies of generated datasets when the conditional form of GAN is employed even in non-private mode. The zone-based accuracy degrades more when the DPCGAN is taken into the account.
It can be also inferred from these two Tables that in the zone-based localization mode, both non-private WGAN and DPWGAN show higher accuracy compared with the non-private CGAN and DPCGAN. 

\begin{table*}	[t]
    \caption{The zone-based accuracy of four classifiers when using original dataset vs generated dataset by non-private WGAN and DPWGAN on $\varepsilon = 1, 5, 10, 15$ and $\delta = 10^{-5}$ (for 384 records). }%
	\makebox[\textwidth][c]{%
		\resizebox{0.85\textwidth}{!}{%
			\begin{tabular}{@{ }l r r r r @{ }}
				\toprule
                & \multicolumn{4}{c}{Accuracy (\%)} \\ \cline{2-5}  
                \multicolumn{1}{l}{} &
				\multicolumn{1}{c}{\textbf{MLP} } &
				\multicolumn{1}{c}{\textbf{Random Forest} } &
				\multicolumn{1}{c}{\textbf{SVM}} &
				\multicolumn{1}{c}{\textbf{Decision Tree}} 
				\\
				\midrule
				\textbf{Original} & 74.0	$\pm$ 3.1  & 90.2 $\pm$ 3.7 & 74.7 $\pm$ 1.1  & 86.7 $\pm$ 2.0  \\	
				\textbf{WGAN (non-private)} & 78.1 $\pm$ 3.4  & 81.5 $\pm$ 4.4 & 83.4 $\pm$ 3.6 & 76.5 $\pm$ 3.2 \\
				\textbf{DPWGAN ($\varepsilon = 15$)} & 77.3 $\pm$ 1.1  & 85.7 $\pm$ 1.5 & 88.9 $\pm$ 0.9 & 75.7 $\pm$ 1.5 \\
                \textbf{DPWGAN ($\varepsilon = 10$)} & 71.3 $\pm$ 2.4 & 71.8 $\pm$ 2.1 & 74.0 $\pm$ 2.1 & 56.6 $\pm$ 3.4 \\	
                \textbf{DPWGAN ($\varepsilon = 5$)} & 63.2 $\pm$ 3.1 & 64.7 $\pm$ 1.9 & 65.2  $\pm$ 2.7 & 56.4 $\pm$ 2.7  \\
                \textbf{DPWGAN ($\varepsilon = 1$)} & 61.7 $\pm$  3.8 & 60.9 $\pm$ 3.2 & 62.5 $\pm$ 1.9 & 62.9 $\pm$ 4.6  \\
				\bottomrule
			\end{tabular}
		}
	}
	
	\label{tab:dpwgan_class}%
\end{table*}

\begin{table*}	[t]
    \caption{The zone-based accuracy of four classifiers when using original dataset vs generated dataset by non-private CGAN and DPCGAN on $\varepsilon = 1, 5, 10, 15$ and $\delta = 10^{-5}$ (for 384 records). }%
	\makebox[\textwidth][c]{%
		\resizebox{0.85\textwidth}{!}{%
			\begin{tabular}{@{ }l r r r r @{ }}
				\toprule
                & \multicolumn{4}{c}{Accuracy (\%)} \\ \cline{2-5}  
                \multicolumn{1}{l}{} &
				\multicolumn{1}{c}{\textbf{MLP} } &
				\multicolumn{1}{c}{\textbf{Random Forest} } &
				\multicolumn{1}{c}{\textbf{SVM}} &
				\multicolumn{1}{c}{\textbf{Decision Tree}} 
				\\
				\midrule
				\textbf{Original} & 74.0	$\pm$ 3.1  & 90.2 $\pm$ 3.7 & 74.7 $\pm$ 1.1  & 86.7 $\pm$ 2.0  \\	
				\textbf{CGAN (non-private)} & 66.2 $\pm$ 3.2  & 65.3 $\pm$ 2.8 & 64.3 $\pm$ 0.7 & 71.1 $\pm$ 0.4 \\
				\textbf{DPCGAN ($\varepsilon = 15$)} & 57.6 $\pm$ 2.4  & 63.9 $\pm$ 1.9 & 61.2 $\pm$ 4.0 & 61.9 $\pm$ 2.2 \\
                \textbf{DPCGAN ($\varepsilon = 10$)} & 53.4 $\pm$ 3.1 & 57.9 $\pm$ 4.5 & 58.8 $\pm$ 1.5 & 57.8 $\pm$ 3.2 \\	
                \textbf{DPCGAN ($\varepsilon = 5$)} & 49.2 $\pm$ 1.5 & 57.6 $\pm$ 3.4 & 51.5  $\pm$ 1.4 & 55.9 $\pm$ 3.5  \\
                \textbf{DPCGAN ($\varepsilon = 1$)} & 46.9 $\pm$  2.7 & 53.8 $\pm$ 2.1 & 49.9 $\pm$ 2.1 & 55.3 $\pm$ 1.2  \\
				\bottomrule
			\end{tabular}
		}
	}
	
	\label{tab:dpcgan_class}%
\end{table*}

\begin{figure}[t]
    \centering
    \includegraphics[width=.49\textwidth]{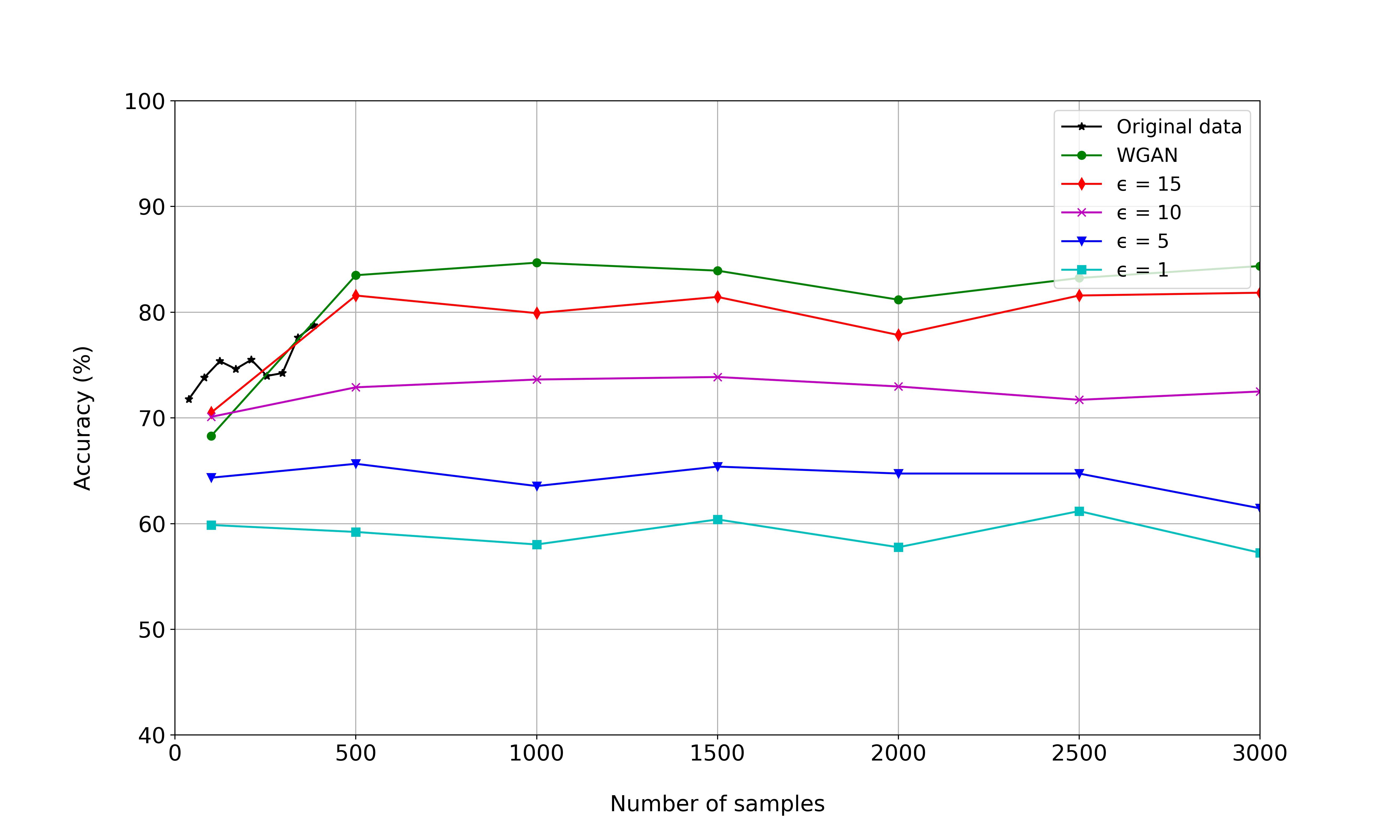}
    \caption{The zone-based accuracy of MLP classifiers when using original dataset vs generated dataset by non-private WGAN and DPWGAN on $\varepsilon = 1, 5, 10, 15$ and $\delta = 10^{-5}$  based on the number of fingerprint samples in the dataset.} 
    \label{fig:dpwgan_class_samples}
\end{figure}

In order to evaluate the accuracy of a zone-based classifier for both DPWGAN and DPCGAN when the number of data points is increased, we consider the MLP classifier as it shows better accuracy in the last two Tables. Figure~\ref{fig:dpwgan_class_samples} illustrates the accuracy of zone-based MLP classifiers when it is trained on the original dataset (dashed black line) vs generated datasets by non-private WGAN and DPWGAN for $\varepsilon = 1, 5, 10, 15$ and $\delta = 10^{-5}$ based on the number of fingerprints in the dataset. As it is mentioned before, the number of records in the original dataset is $384$, while for the synthetic datasets, we tried to generate more fingerprint records so that we could improve the localization accuracy as well. The results show that higher accuracy occurs in the original, non-private WGAN and DPWGAN with $\varepsilon = 15$ datasets when the data points incremented up to 500; however, for more private datasets, higher number of records has no specific influence on the accuracy.

\begin{figure}[t]
    \centering
    \includegraphics[width=.49\textwidth]{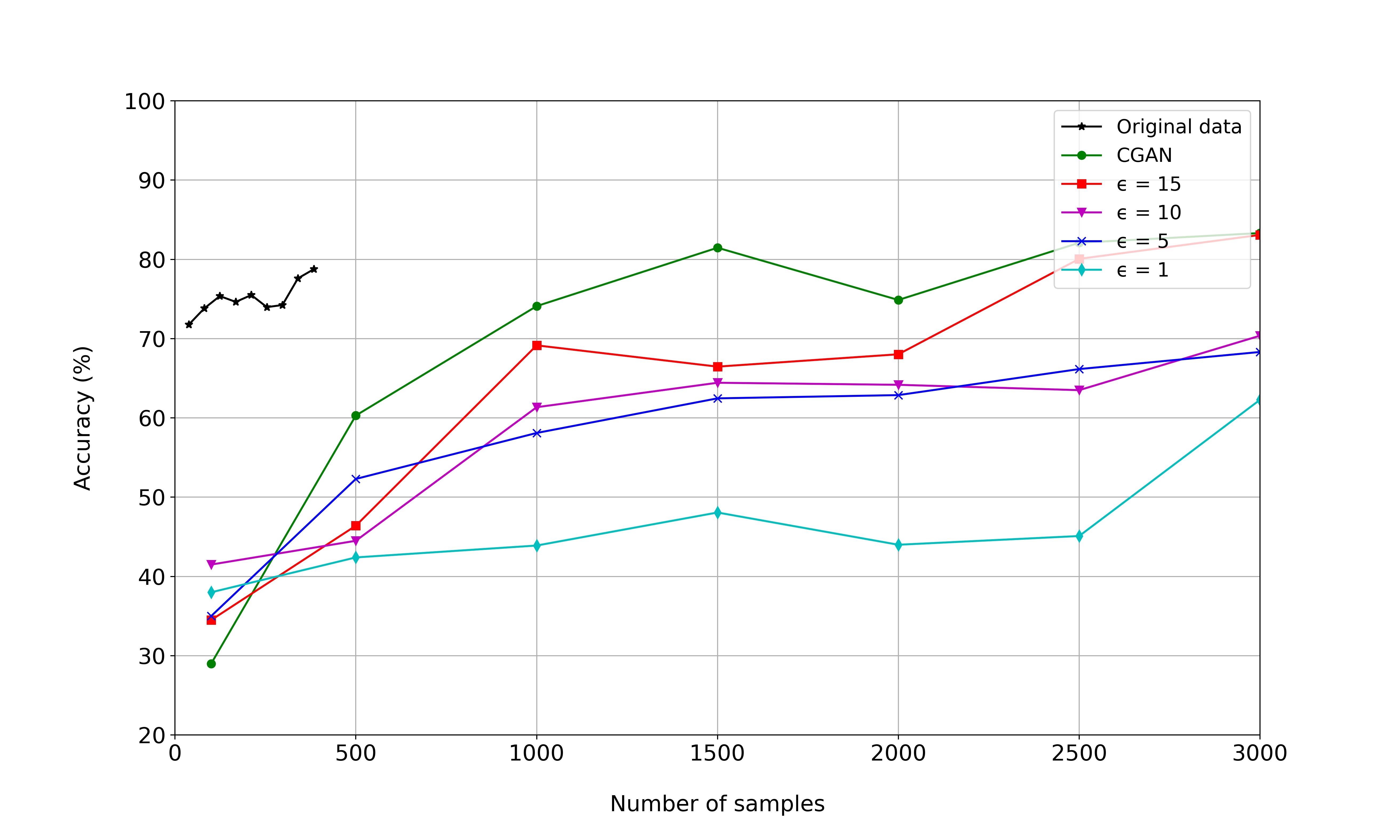}
    \caption{The zone-based accuracy of MLP classifiers when using original dataset vs generated dataset by non-private CGAN and DPCGAN on $\varepsilon =1, 5, 10, 15$ and $\delta = 10^{-5}$  based on the number of fingerprint samples in the dataset.} 
    \label{fig:dpcgan_class_samples}
\end{figure}

Besides, the accuracy of zone-based MLP classifiers for CGAN and DPCGAN is plotted in Figure~\ref{fig:dpcgan_class_samples}. Here also the classifier is trained on the original dataset (dashed black line) vs generated datasets by non-private WGAN and DPWGAN for $\varepsilon = 1, 5, 10, 15$ and $\delta = 10^{-5}$ based on the number of data points in the dataset.
It can be seen that the zone-based accuracies of the trained classifier by non-private CGAN and DPCGAN are much lower than the accuracy of the trained classifier by the original dataset. However, it has been improved when it has learned from more synthetic records. The reason for this is that there are $15$ labels for all $384$ records in a zone-based mode, so we have a low number of records for each label which is not that helpful for training the classifiers. With a higher number of records, on the other hand, more data is involved to train the synthetic dataset for every label.

In addition, note that since DPCGAN is conditionally trained based on the labels, the higher number of records affects this conditionally private GAN more in comparison with DPWGAN, which labels are also generated similarly to RSS features. It is also observed that for a higher number of records, even more private datasets achieve the same or higher level of accuracy than the original dataset.

\subsection{Privacy Evaluation}
To measure the disclosure risk of our model, we compute the average Euclidean distances between the generated data points and the closest point in the original dataset. 
Formally, let $D$ and $D'$ denote the original and generated dataset, respectively. The disclosure risk is then defined as follows:
\begin{align}
    & Disclosure_{Min}(D, D') =  \frac{1}{ |D'| } \sum_{x' \in D'}  \argmin_{x \in D}  \lVert x-x'  \lVert 
\end{align}
where $|D'|$ denotes the number of records in $D'$, and $ \lVert \cdot  \lVert $ is the Euclidean distance between the points $x \in D$ and $x' \in D'$; $\argmin$ selects the minimum of these distances, and  $\frac{1}{ |D'| } $ returns the average of the associated distances. It should be noted that the higher value of $ Disclosure_{Min}$ means that the disclosure risk (privacy violation) is lowered.

Table~\ref{tab:disclosure} reports the disclosure risk (in meters) for the generated dataset linked to the original dataset. It can be observed that, as expected, when more privacy is obtained ($\varepsilon$ reduces), the disclosure risk is reduced ($Disclosure_{Min}$ increases).  
In general, among the three techniques, the non-private CGAN and DPCGAN show higher privacy gain (higher $Disclosure_{Min}$) in the zone-based mode, and non-private WGAn and DPWGAN show lower privacy gain (lower $Disclosure_{Min}$) in the location-based mode.

\begin{table*}[!t]
\caption{Average minimum Euclidean distances.}
\begin{subtable}{1\textwidth}
\sisetup{table-format=-1.2}  
\resizebox{1\textwidth}{!}{
 \begin{tabular}{|c|*{5}{c|}}
        \hline
         & \multicolumn{5}{c|}{ \textbf{Location-based (DP)WGAN} }  \\
        \cline{2-6}
        & \textbf{WGAN (non-private)} & \textbf{DPWGAN ($\varepsilon=$15)} & \textbf{DPWGAN ($\varepsilon=$10)} & \textbf{DPWGAN ($\varepsilon=$5)} & \textbf{DPWGAN ($\varepsilon=$1)} \\
        \hline
        $Disclosure_{Min} (m)$ & 5.88 & 6.20 & 6.56 & 6.98 & 7.01\\
        \hline
    \end{tabular}
}
 \caption{Disclosure risk of the dataset generated by non-private WGAN and DPWGAN in the location-based localization.}
    \label{tab:risklocationw}
\end{subtable}
\begin{subtable}{1\textwidth}
\sisetup{table-format=-1.2}  
\resizebox{1\textwidth}{!}{
 \begin{tabular}{|c|*{5}{c|}}
        \hline
         & \multicolumn{5}{c|}{ \textbf{Zone-based (DP)WGAN} } \\
        \cline{2-6}
        & \textbf{WGAN (non-private) } & \textbf{DPWGAN ($\varepsilon=$15)} & \textbf{DPWGAN ($\varepsilon=$10)} & \textbf{DPWGAN ($\varepsilon=$5)} & \textbf{DPWGAN ($\varepsilon=$1)} \\
        \hline
        $Disclosure_{Min} (m)$ & 6.78 & 7.54 & 8.03 & 8.81 & 9.89 \\
        \hline
    \end{tabular}
}
 \caption{Disclosure risk of the dataset generated by non-private WGAN and DPWGAN in the zone-based localization.}
    \label{tab:riskzonew}
\end{subtable}
\begin{subtable}{1\textwidth}
\sisetup{table-format=-1.2}  
\resizebox{1\textwidth}{!}{
 \begin{tabular}{|c|*{5}{c|}}
        \hline
         & \multicolumn{5}{c|}{\textbf{Zone-based (DP)CGAN} }  \\
        \cline{2-6}
        & \textbf{CGAN (non-private)}& \textbf{DPCGAN ($\varepsilon=$15)} & \textbf{DPCGAN ($\varepsilon=$10)} & \textbf{DPCGAN ($\varepsilon=$5)} & \textbf{DPCGAN ($\varepsilon=$1)} \\
        \hline
        $Disclosure_{Min} (m)$ & 7.50 & 7.92 & 11.38 & 14.68 & 18.38 \\
        \hline
    \end{tabular}
}
 \caption{Disclosure risk of the dataset generated by non-private CGAN and DPCGAN in the zone-based localization.}
    \label{tab:riskzonec}
\end{subtable}
\vspace{-0.5cm}
\label{tab:disclosure}
\end{table*}

\subsection{Discussion}
The amount of noise that is added to the gradients can be controlled by the privacy budget ($\varepsilon$) and the choice of the DP mechanism used, which can be optimized to balance utility and privacy. Therefore, while the privacy budget (controlled by $\varepsilon$) does affect the privacy guarantees of DPGANs, it has a relatively small impact on the quality of the generated data compared to other factors. This is because the DP mechanism is only applied to the gradients of the Generator, which are used to update the parameters of the Generator during the training process. The generated data is not directly affected by the DP mechanism, as it is only indirectly affected by the gradients. The choice of $\varepsilon$ should be then made carefully to balance utility and privacy, but other factors such as the architecture of the GAN, the quality of the training data, and the choice of hyper-parameters are also important for the quality of the generated data.

Furthermore, our experimental results for the DPGAN framework on indoor location data suggest using DPWGAN in location-based localization, whenever accuracy is a matter of importance. However, this accuracy comes with the price of privacy loss. On the other hand, DPCGAN in zone-based mode shows higher privacy gain with lower accuracy, and DPWGAN in zone-based mode shows the best trade-off between privacy gain and utility loss. These results provide insight into the selection of the appropriate DPGAN model for the indoor location framework based on the need of users, which is higher utility, higher privacy, or a better trade-off between utility and privacy.

\section{Conclusion and Future Work}\label{sec:conclusion}
In this paper, we suggest an indoor location framework based on differentially private GANs (DPGANs) aiming to generate a privacy-preserving synthetic dataset for both location-based and zone-based indoor localization systems.
We specifically investigate the use of two popular DPGANs, namely DPWGAN and DPCGAN. DPGANs can provide a promising approach for balancing the competing demands of privacy and accuracy by generating synthetic location data that is statistically similar to the real data but does not reveal specific information about any individual user. This can enable accurate location-based services to be provided without compromising user privacy. 
We have also exhibited that synthetic data can enhance the performance of fingerprinting localization, especially in situations where the procedure of collecting data is expensive and time-consuming.
While the DPGANs framework can ensure the confidentiality of training data, the synthetic data is vulnerable in very private modes especially when the privacy budget is less than $1$. Therefore, improving the indoor location DPGAN framework for higher privacy is one of the future works. In addition, the current framework is limited by having only a single Generator and Discriminator. Hence, a promising avenue for further investigation could be to expand this framework to encompass multiple Generators and Discriminators, enabling them to tackle more intricate problems.

\section{Declarations}\label{sec:dec}
\subsection{Code availability}
The dataset used and analysed during the current study is available from the corresponding author upon reasonable request.

\subsection{Competing interests}
The authors have no competing interests to declare relevant to this article's content.

\subsection{Ethics approval}
The authors have no relevant financial or non-financial interests to disclose, and no funding was received for conducting this study. In addition, no human participants or animals are involved in this research. 

\bibliographystyle{IEEEtran}
\bibliography{template}

\end{document}